\documentclass[aps,twocolumn,pra,10pt]{revtex4-1}
\usepackage{graphics}
\usepackage{helvet}
\usepackage{dcolumn}
\usepackage{bm}
\usepackage{amsmath}
\usepackage{hyperref}
\hypersetup{colorlinks=true, urlcolor=blue}
\usepackage{color}
\usepackage{amssymb}
\usepackage{amsfonts}
\usepackage{graphicx}
\usepackage{physics}
\usepackage{amsthm}
\usepackage{float}
\usepackage{subfigure}

\begin{document}

\title{Robustness of interferometric complementarity under Decoherence}

\author{Gautam Sharma}
\affiliation{Harish-Chandra Research Institute, HBNI, Chhatnag Road, Jhunsi, Allahabad 211 019, India}

\author{Mohd Asad Siddiqui}
\affiliation{Harish-Chandra Research Institute, HBNI, Chhatnag Road, Jhunsi, Allahabad 211 019, India}

\author{Shiladitya Mal}
\affiliation{Harish-Chandra Research Institute, HBNI, Chhatnag Road, Jhunsi, Allahabad 211 019, India}

\author{Sk Sazim}
\affiliation{Harish-Chandra Research Institute, HBNI, Chhatnag Road, Jhunsi, Allahabad 211 019, India}

\author{Aditi Sen(De)}
\affiliation{Harish-Chandra Research Institute, HBNI, Chhatnag Road, Jhunsi, Allahabad 211 019, India}

\begin{abstract}
    Interferometric complementarity is known to be one of the most nonclassical manifestations of the quantum formalism. It is commonly known as wave-particle duality and has been studied presently from the perspective of quantum information theory where wave and particle nature of a quantum system, called quanton, are characterised by coherence and path distinguishability respectively. We here consider the effect of noisy detectors on the complementarity relation. We report that by suitably choosing the initial quanton and the detector states along with the proper interactions between the quanton and the detectors, one can reduce the influence of noisy environment on complementarity, thereby pushing it towards saturation. To demonstrate this,  three kinds of noise on detectors and their roles on the saturation of the complementarity relation are extensively studied. We also observe that for fixed values of parameters involved in the process, asymmetric quanton state posses low value of coherence while it can have a higher amount of distinguishability, and hence it has the potential to enhance the duality relation. 
\end{abstract}

\maketitle

\section{Introduction}
The wave-particle duality or interferometric complementarity  \cite{ein05,comp23,brog24}, exhibiting both wave- and particle-like behaviour of quantum systems which are often called quantons  \cite{bunge,levy}, plays a significant role in quantum mechanics. In 1928, Bohr first  pointed out that this wave and particle nature of quantum systems are ``exclusive" to each other \cite{bohr}. Later on, an  information-theoretic notion was used to obtain complementarity relation  by  Wootters and Zurek which qualitatively shows the impossibility of simultaneously observing both path information and fringe visibility in an interference experiment \cite{wootters}. A quantitative version of this relation was proposed by Greenberger and Yasin which involve predictability, defined as the difference between probabilities of going through two different paths by the initial quanton, and the visibility of an interference pattern, denoted by \( \mathcal{V}\) \cite{greenberger}. 
To gather information about the path, Englert had introduced detectors which can distinguish which-way the quantum systems travel. Such a posteriori path information acquired through measurement, denoted as  $\mathcal{D}$, revealing the particleness of quantum systems and the visibility, admitting the wave nature of quantum systems, lead to a duality relation, given by \cite{prl96}
\begin{equation}
\mathcal{D}^2 + \mathcal{V}^2 \leq 1.
\label{englert}
\end{equation}

In the past two decades, a lot of effort has been put towards addressing several questions on the relations involving various interference set-ups, source with single or multiphotons  \cite{durr,mei,luis,bimonte,bimonte1,li2012,englertmb,banaszek,3slit,cd,3slitn,nonlocal,bagan,coles1,biswas,tqma,menon,predict,misba,qur19,Zubairy} etc. 
Moreover, it has also been experimentally tested in various kinds of physical systems including atom \cite{Durr'98}, nuclear magnetic resonance \cite{Peng03}, single-photon source \cite{Jacques'08}.

On the other hand,  Heisenberg's uncertainty principle limits the precision of outcome-statistics of two complementary observables \cite{Heisen'27}. Recent developments show that it is important in applications of quantum information theory like detection of entanglement \cite{Guhne'04}, quantum steering \cite{Walborn'09},  mixedness of quantum systems \cite{mal'13}. Initially, uncertainty relation was considered as the quantitative version of the complementarity relation between wave and particle nature of quantum states \cite{Bohr'49}. However, such connection originates lots of debates \cite{mirWiseman07}.   Invoking a scenario, it was claimed  \cite{scully, prl96} that interferometric complementarity is not connected with uncertainty. Later on it was argued that in every interferometer  scenario, one can derive an uncertainty relation of some observables \cite{Storey'94, Wiseman'97} (cf.  \cite{Qureshi'18, Busch'06}). 

In the literature,   path distinguishability is quantified in two different ways -- (1) minimum error state discrimination (MESD), established by Helstrom \cite{Helstrom'76}, where in each run of the experiment,  one has to guess the input state with least probability of error \cite{Bergou'10}  and (2)  unambiguous state discrimination protocol, in which the state has to be identified always successfully, minimising the probability of inconclusive case \cite{uqsd}. 
On the other hand, in recent years to characterise the wave property, instead of the fringe visibility,  coherence  \cite{coherence}, $\mathcal{C}$, of the reduced quanton state after interaction with the detectors is prescribed. By considering the probability of success for distinguishing states detected in different paths, denoted by $\mathcal{D}_Q$ and the coherence, the new linear complementarity relation of the form  $\mathcal{D}_Q+\mathcal{C}\le 1$ was derived \cite{cd} (see also \cite{tqma}). In a similar spirit of the inequality (\ref{englert}), Bagan \emph{et al.} \cite{bagan} found a quadratic complementarity relation where wave nature is captured by coherence and distinguishability is measured by the success probability of state discrimination protocol \cite{Helstrom'76}.

In most of the cases, interferometric complementarity, a genuine manifestation of quantumness, was studied in an ideal experimental set-up. However, a noisy environment is unavoidable in any experiment with quantum systems.  Therefore, studying the effects of noise on the duality relation is interesting although investigations in this direction are limited (c.f. \cite{jia14, wang16}). In this paper,  we consider a situation, where the initial detector states are affected by noise.  We then investigate its consequences on the quadratic complementarity relation derived in Ref. \cite{tqma}. Specifically,  we address the question - how is the complementarity modified if path information is obtained through damped detectors in the context of double-slit experiment \cite{prl96, tqma}. Towards answering this,  three kinds of noisy channels, namely, depolarizing (DC), amplitude damping (ADC) and phase damping noises (PDC) are considered and in presence of all the noisy channels, we observe that in general, the complementarity goes far from saturation. However, we find that the decohering effect can be countered by monitoring the relevant parameters involved in the set-up, such as the initial detector state, the initial quanton and the interaction between the quanton and the  detectors. In the case of the depolarizing channel, for symmetric quantons, we find that although the path distinguishability and the coherence independently change with the initial detector states, duality relation is independent of the initial detector parameters.  
We notice that in the presence of strong noisy environment, there exists a finite region in parameter settings for which complementarity relation can be saturated by proper controlling of the system parameters. We also observe that the asymmetric initial quanton state, with suitably chosen detector state, can facilitate the distinguishability and hence duality relations in a non-trivial way. The advantage of asymmetry over symmetric quanton from the perspective of saturation of the bound is reported for all the considered noisy channels.  

The paper is organized as follows: In Sec. \ref{Preliminaries}, we describe the interferometric set-up required for the duality test in a noisy environment, we discuss all the definitions used and the complementary relation. The consequences of noisy channels on complementarity  are discussed in Sec. \ref{sec_noise}.  Specifically, Subsecs. \ref{subsec:depol}, \ref{subsec_ampd}, \ref{subsec_phasedamp} deal with the depolarizing,  the amplitude damping and the phase damping channels respectively. The Kraus operators for noisy channels are discussed in the Appendix. We finally conclude in Sec. \ref{sec_conclu}.

\section{Complementarity relation between distinguishability and coherence}
\label{Preliminaries}
In this section, we set the stage for studying complementarity relation in terms of path distinguishability and coherence in the presence of noise. Before presenting the complementarity relation, we first discuss  the picture considered here for the duality test and then give the definitions of distinguishability and coherence considered in this paper. 

\subsection{Interferometeric set-up in  noisy scenario}
\begin{figure}[h]
	\includegraphics[width=0.5\textwidth, keepaspectratio]{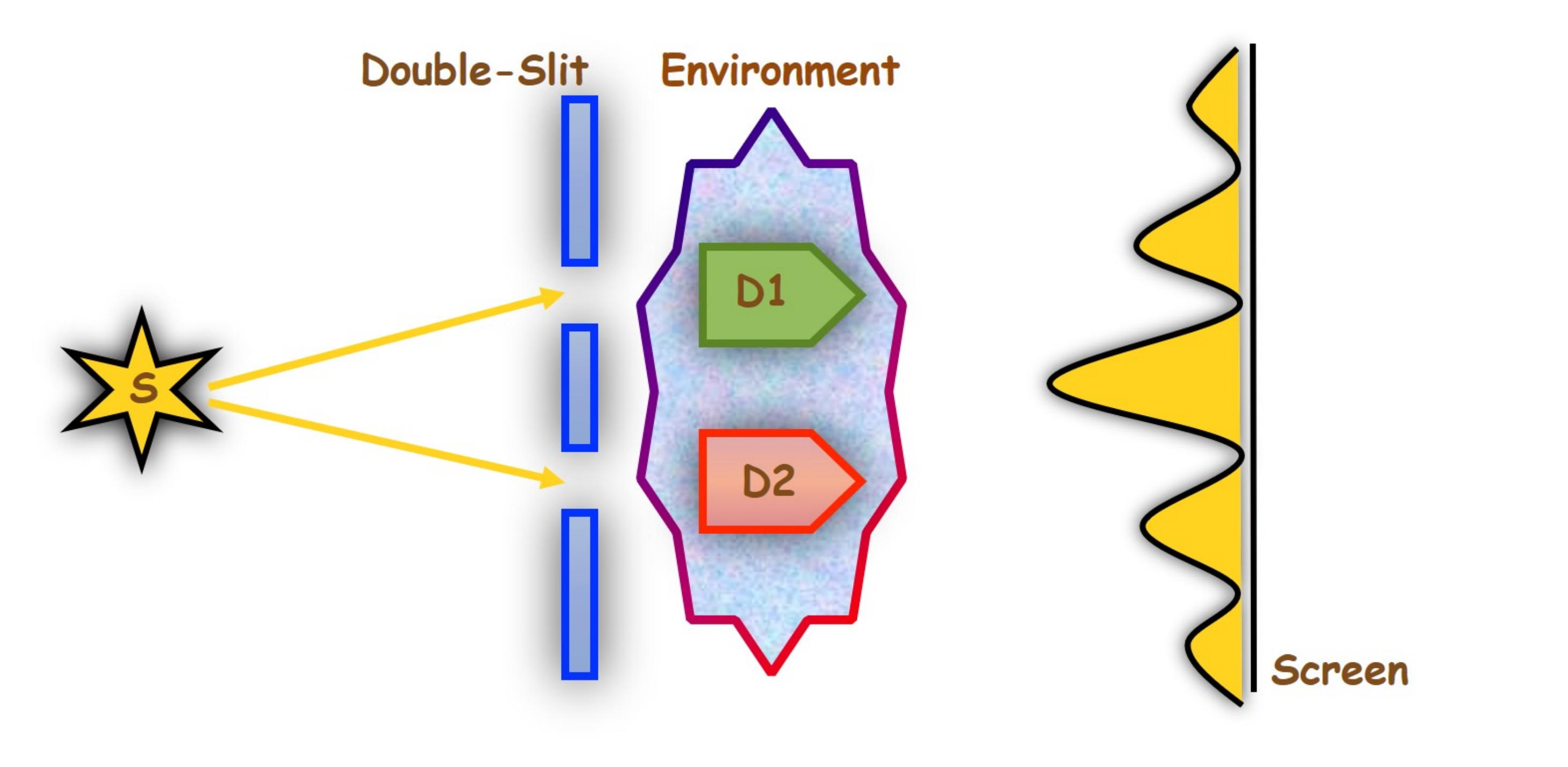} 
	\caption{(Color online): A schematic diagram of the double-slit interference experiment.  The detector which is used to distinguish the path taken by the quanton is affected by noise shown in bluish gray color. }\label{schematic}
\end{figure}

We consider a double-slit interference experiment, with pure initial quanton state, $|\psi_{in}\rangle=\sum_{i=1}^2{\sqrt{p_i}}{|\psi_i\rangle}$, where $\{|\psi_1\rangle,|\psi_2\rangle\}$ forms an orthonormal basis,  with ${p_i}$ being the probability of acquiring the path, $i$. Initially the joint state of a quanton and a detector system is given by $\rho^{\rm(in)}_{sd}=\rho_{in}\otimes \rho^{(0)}_d$ (see Fig. \ref{schematic}). 

In presence of noise, the detector state gets modified, as
\begin{equation}
\label{rhon}
\rho_d^{(0)} \longrightarrow \widetilde\rho_d^{(0)} =\sum_{i}K_{i}\rho_d^{(0)} K^{\dagger }_{i},
\end{equation}
with completeness relation $\sum_{i}K^{\dagger }_{i}K_{i}=\mathcal{I}$. The spectral decomposition of  the transformed detector state then can be written as
\begin{equation}
\label{spectral}
\widetilde\rho_d^{(0)} =\sum_{k} {D_{k}} |{d_k}\rangle \langle{d_k}|,
\end{equation}  
where  $\sum_{k} D_k = 1$, $D_k\ge0$, and $ \langle d_k|d_l \rangle =\delta_{kl}$.

While acquiring the first or the second path,  the detector state gets transformed as
\begin{eqnarray}
\label{4}
\rho^{(i)}_d=U^\dag_{i} \widetilde\rho_d^{(0)} U_{i},\ \ i=1,2,
\end{eqnarray}
where $\{U_{i}\}$  denotes the set of unitary transformations acted on the detector state, corresponding to the path of the quanton. 

The global interaction creates entanglement between the detector and the quanton. In general, the controlled unitary operations can be used to correlate the quanton and the detector. Thus, for gaining the knowledge of the quanton, it is sufficient to make a quantum measurement on the detector state. The combined quanton-detector state can now be written as

\begin{equation}
\label{5}
\rho_{sd}=\sum_{i,j=1}^2\sqrt{p_i} \sqrt{p_j}\, |\psi_i\rangle \langle \psi_j|\otimes U_i^\dag \widetilde\rho_d^{(0)} U_j.
\end{equation}
Given this kind of interaction, we measure the coherence of the quanton and path distinguishability in this situation to estimate distinguishability-coherence complementarity. \\

\textit{\textbf{Quantum coherence}}: Quantum coherence \cite{coherence} captures the wave nature of a quanton, defined as
\begin{equation}
	\label{1}
	{\mathcal C}(\rho) = {1\over n-1}\sum_{i\neq j} \abs{\rho_{ij}} ,
\end{equation}
where $n$ is the dimensionality of the Hilbert space. For $n=2$, $ {\mathcal C} = \abs{\rho_{12}}+ \abs{\rho_{21}}. $
Note here that although coherence is a property of a quanton state alone, due to the entanglement between the detector and the quanton, noise on detector affects coherence as well. Therefore, in this scenario, we first find the reduced density matrix of the quanton, by tracing over the detector states, given by
\begin{eqnarray}
\label{6}
\rho_s &=& \sum_{i,j=1}^2 \sqrt{p_i} \sqrt{p_j} \ \mbox{Tr}\left( U_i^\dag \widetilde\rho^{(0)}_d U_j \right) \ |\psi_i\rangle\langle\psi_j|.
\end{eqnarray}

For the given interferometric set-up, the set $\{ |\psi_i\rangle\}$ forms the complete incoherent basis. The  coherence in this basis now can be calculated for the quanton using the reduced density matrix, as
\begin{eqnarray}
\label{7a}
{\mathcal C} &=& \sum_{i\neq j} \abs{\langle\psi_i|\rho_s|\psi_j\rangle}
\nonumber\\
&=& \sum_{i\neq j} \sqrt{p_i} \sqrt{p_j}  \ \abs{ \mbox{Tr}\left( U_i^\dag \widetilde\rho^{(0)}_d U_j \right) }. \nonumber
\end{eqnarray}
Using Eq. (\ref{spectral}), we get the following form:
\begin{eqnarray}
\label{7}
{\mathcal C}
&=&  2\sqrt{p_1} \sqrt{p_2}  \ \abs{\sum_k {D_k} \ \langle {d_k}|U_2 U_1^\dag  |{d_k}\rangle}.
\end{eqnarray}

\textit{\textbf{Path distinguishability}}: 
The path information of quanton states can be obtained by distinguishing the given detector states $\{\rho^{(1)}_d, \rho^{(2)}_d\}$, which appear with probabilities $\{p_{1}, p_{2}\}$. 
For  distinguishing two non-orthogonal states,  there exist  measurement strategies which give an upper bound on the success probability of discrimination.
We will use the upper bound of success probability based on MESD, as introduced by Bagan et. al. \cite{bagan}, given by
\begin{eqnarray}
	\label{Ps}
	P_s \le \frac{1}{2}+ \frac{1}{4}  \sum_{i,j=1}^{2} \abs{\abs{ p_{i}\, \rho^{(i)}_d- p_{j}\, \rho^{(j)}_d}},
\end{eqnarray}
where $\abs{\abs{\, X\,}}={\rm Tr}\,[\sqrt{X^\dag\,X}]$.

The path distinguishability  characterizes the particle nature of the quanton and its quantifier can be written as  \cite{prl96, Bergou'10, li2012}
\begin{eqnarray}
	\label{dis}
	{\cal D}= 2 P_s-1 \le \abs{\abs{ p_{1}\, \rho^{(1)}_d- p_{2}\, \rho^{(2)}_d}}.
\end{eqnarray} 

The path quantifiers have also been proposed from unambiguous quantum state discrimination(UQSD) method \cite{cd, tqma}. In this paper, we will stick to MESD to quantify the particleness.

Using Eqs. (\ref{spectral}) and (\ref{4}), the above Eq. (\ref{dis}) reduces to 
\begin{eqnarray}
	{\cal D}\le \sum_k D_k \ (1-4 p_1 p_2\ \vert \langle d_k| U_2  U_1^\dag |d_k\rangle \vert ^2)^{1/2}.
	\label{Dqnew}
\end{eqnarray} 

For symmetric quanton state $(p_1 =p_2)$, $0\leq {\cal D}\leq 1$, where perfect distinguishability occurs with ${\cal D} = 1$, obtained for orthogonal states. For non-orthogonal $\{\rho^{(i)}_d$\}'s, $0 \le {\mathcal D} < 1$.
If the  quanton state is asymmetric, i.e.,  the $(p_1 \ne p_2)$ then,  ${\cal P} \leq {\cal D} \leq 1$ \cite{prl96}, where ${\cal P}= \abs{p_1-p_2}$, is  known as predictability \cite{greenberger}.

\subsection{Complementarity  Relation}\label{sec:CR}
By incorporating the effect of noise in the interferometric set-up, we have the expressions of coherence and path distinguishability in Eqs. (\ref{7}) and (\ref{Dqnew}) respectively. By squaring and adding them we get the desired complementarity relation as
\begin{equation}
	\mathcal{F}  \equiv {\mathcal C}^2 + {\mathcal D}^2 \le \sum_{k,l} {D_k} {D_l}=\abs{\abs{ \widetilde\rho^{(0)}_{d}}}^2=1,
	\label{fduality}
\end{equation}
where we call \(\mathcal{F}\) as the complementarity function. 
Eq. (\ref{fduality}) represents the duality relation for the double-slit interference experiment. Note that when both the quanton and the detector states are pure, $\mathcal{F}=1$.  We will show that with a proper choice of system parameters, how complementarity relation can reach saturation even for noisy detector states.

\section{Analysis of noisy detector in double-slit experiment}
\label{sec_noise}

We are now ready to investigate the situation where the test for the wave-particle duality relation is disturbed by three paradigmatic noise models, acted on the detector states. We will subsequently show that by suitably tuning the parameters involved in the process, one can suppress the decoherence in a certain way.

 Let us illustrate the double-slit scenario described in Sec. \ref{Preliminaries}. The initial quanton and the detector states are taken respectively as

\begin{eqnarray}
|\psi_{in}\rangle & = & \sqrt{p_1}  |\psi_{1}\rangle + \sqrt{1-p_1} |\psi_{2}\rangle, \nonumber \\
|d_{0}\rangle &= &  \cos{\frac{\theta}{2}}\, |\psi_{1}\rangle + e^{i \phi} \sin{\frac{\theta}{2}}\, |\psi_{2}\rangle,
\label{eq:initial}
\end{eqnarray}
where \(\langle\psi_{1}| \psi_{2}\rangle =0\),  $\phi \in [0,2\pi]$ and $\theta \in [0,\pi]$. 

We now assume that before interacting with the quanton, the detector state gets modified under the noise present in the system, which can be represented by Eq. (\ref{rhon}). Then the controlled unitary operations correlate the initial quanton and the modified path detector. For our purpose, we take the following form of unitaries: 

\begin{equation}
U_1 =\left(
\begin{array}{cc}
e^{i (\text{$\alpha_1$}-\frac{\text{$\beta_1$}}{2}-\frac{\text{$\delta_1 $}}{2})} \cos \frac{\text{$\eta_1$}}{2} & -e^{i \left(\text{$\alpha_1 $}-\frac{\text{$\beta_1 $}}{2}+\frac{\text{$\delta_1 $}}{2}\right)} \sin \frac{\text{$\eta_1 $}}{2} \\
e^{i \left(\text{$\alpha_1 $}+\frac{\text{$\beta_1 $}}{2}-\frac{\text{$\delta_1 $}}{2}\right)} \sin \frac{\text{$\eta_1 $}}{2} & e^{i \left(\text{$\alpha_1 $}+\frac{\text{$\beta_1 $}}{2}+\frac{\text{$\delta_1 $}}{2}\right)} \cos \frac{\text{$\eta_1 $}}{2} \\
\end{array}
\right),
\label{eq:unitary1}
\end{equation}
and 
\begin{equation}
U_2 =\left(
\begin{array}{cc}
e^{i (\text{$\alpha_2$}-\frac{\text{$\beta_2$}}{2}-\frac{\text{$\delta_2 $}}{2})} \cos \frac{\text{$\eta_2$}}{2} & -e^{i \left(\text{$\alpha_2 $}-\frac{\text{$\beta_2 $}}{2}+\frac{\text{$\delta_2 $}}{2}\right)} \sin \frac{\text{$\eta_2 $}}{2} \\
e^{i \left(\text{$\alpha_2 $}+\frac{\text{$\beta $}}{2}-\frac{\text{$\delta_2 $}}{2}\right)} \sin \frac{\text{$\eta_2 $}}{2} & e^{i \left(\text{$\alpha_2 $}+\frac{\text{$\beta_2 $}}{2}+\frac{\text{$\delta_2 $}}{2}\right)} \cos \frac{\text{$\eta_2 $}}{2} \\
\end{array}
\label{eq:unitary2}
\right),
\end{equation}
where  $\alpha_i$, $\beta_i$, $\delta_i$, and $\eta_i$ \((i =1, 2)\) are real numbers. Note that one of the phases, $\alpha_i$s,  of $U_i$s do not contribute in the computation of $\mathcal{C}$ and $\mathcal{D}$ since the moduli of the unitaries are present in the expressions for both the cases.
Therefore, $ \mathcal{F}$ is a function of all these parameters including the noise parameter $\gamma$, defined in Appendix and should satisfy the complementarity i.e.,
\begin{equation}
\mathcal{F} (\beta_1,\beta_2, \delta_1,\delta_2, \eta_1,\eta_2, \theta,\phi, p_1, \gamma) \le 1.
\label{func}
\end{equation}
Given a fixed value of $\gamma$, we aim to control all these parameters suitably, so that $\mathcal{F}$ saturates or goes close to a saturation value. In particular, for a fixed noise model, we analyse the pattern of $\mathcal{F}$ systematically by fixing some parameters to a certain fixed value and varying the rest of the parameters.

\subsection{Depolarizing channel}
\label{subsec:depol}

\begin{figure}[H]	
	
	\subfigure[]{
		\includegraphics[width=0.227\textwidth, height=0.17\textwidth]{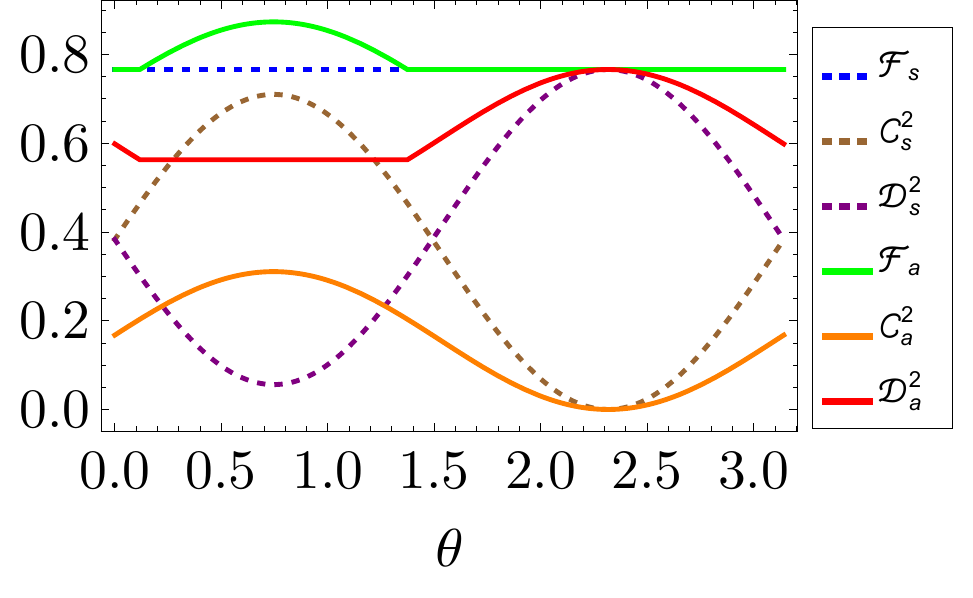}       
		\label{fig6}
	}
	\subfigure[]{
		\includegraphics[width=0.227\textwidth, height=0.17\textwidth]{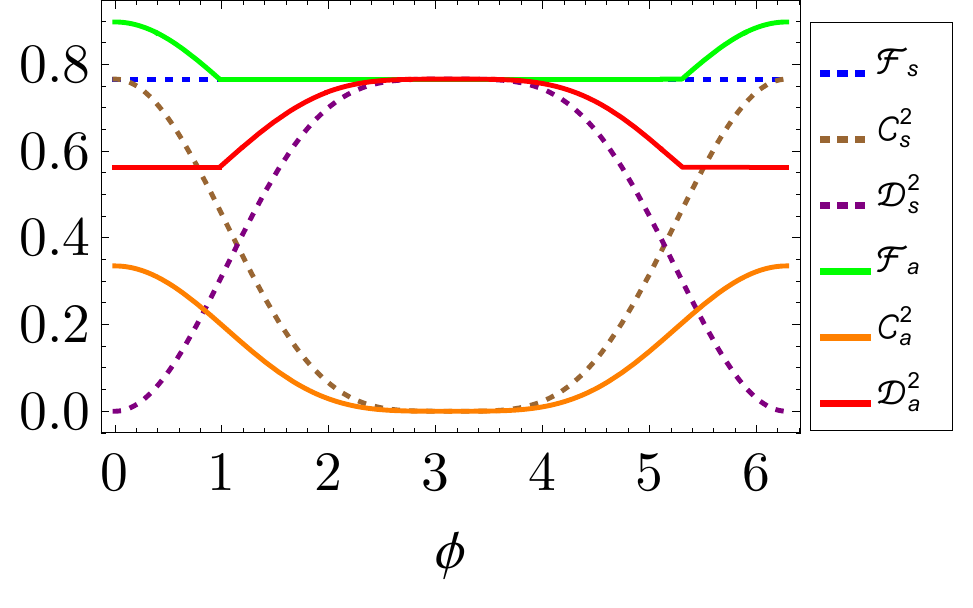}
		\label{fig7}
	}
	\caption{(Color online) Plots of coherence, distinguishability and complementarity function for the depolarizing channel. The horizontal axes represent  $\theta$ in (a) and $\phi$ in (b). 
For the symmetric initial quanton for which quantities are marked with subscript `s', $\mathcal{C}_s^2$, $\mathcal{D}_s^2$ clearly depend on $\theta$ and $\phi$ while $\mathcal{F}_s$ remains independent of both. However, in the asymmetric case, with $p_1=\frac{1}{8}$,  $\mathcal{F}_a$ depends on \(\theta\) and $\phi$. Clearly, for certain range of \(\theta\), \(\mathcal{F}_a > \mathcal{F}_s\).  We fix  $\text{$\eta_1$}=\pi ;\text{$\eta_2$}=\frac{\pi }{2};\text{$ \Delta \beta $}=2 \pi ;\text{$\Delta \delta $}=\pi; \gamma =\frac{1}{8}$ in both the figures. In (a):  $\phi =\frac{\pi }{8}$ while in (b):  $\theta =\frac{\pi }{4}$.  }
	\label{f_distr}
\end{figure}

Let us first consider the depolarizing channel, described in Appendix \ref{subsec:depolK}. It is clear from Eq. (\ref{21a}) that the state
symmetrically changes its form after sending it through a  DC. 
We start analyzing its effect with the symmetric initial pure state, i.e., \(|\psi_{in}\rangle = \frac{1}{\sqrt{2}}(|\psi_{1}\rangle + |\psi_{2}\rangle)\). In this situation, the left-hand side of (\ref{func}) takes the form as
\begin{widetext}
\begin{align}
\label{dualitydepol}
\mathcal{F}=& \frac{1}{2}\left(\big(2+(1-\cos(\Delta \beta+\Delta \delta) (\gamma-2)\gamma \big) \cos^2\frac{\eta_1}{2}\cos^2\frac{\eta_2}{2}+ (2+(1-\cos(\Delta \beta-\Delta \delta)) (\gamma-2)\gamma \big) \sin^2\frac{\eta_1}{2}\sin^2\frac{\eta_2}{2} \right) \nonumber \\  \nonumber\\ 
&+\frac{1}{2}(\gamma-1)^2(1-\cos\eta_1\cos\eta_2)-\frac{1}{4}(\gamma-2)\gamma\big[(\cos\Delta \beta+\cos\Delta\delta)\sin\eta_1\sin\eta_2\big],
\end{align}
\end{widetext}	
where $\Delta \beta=\beta_1-\beta_2$ and $\Delta \delta=\delta_1-\delta_2$. Notice first that the expression in  Eq. (\ref{dualitydepol}) is independent of both the parameters in detectors,  $\theta$ and $\phi$, which implies that the manipulation of the initial detector state can not be useful to overcome the aftermath of the depolarizing noise.
Although $\mathcal{F}$ is independent of $\theta$ and $\phi$, coherence and path distinguishability depends individually on the $|\psi_{in}\rangle $ as depicted in Figs. \ref{fig6} and \ref{fig7}. \\

We first notice that the expression for $\mathcal F$ depends  on the differences in phases of \(U_1\) and \(U_2\), i.e., on $\Delta \beta$ and $\Delta\delta$ as well as the trigonometric angles, \(\eta_1\) and $\eta_2$. Hence, to diminish the power of \(\gamma\) on \(\mathcal{F}\), the parameters that can be controlled are $\Delta \beta$ and $\Delta \delta$, $\eta_1$, and $\eta_2$.  Let us investigate the patterns of  \(\mathcal{F}\) for different values of these parameters whose functional dependence on $\mathcal{F}$ is similar.

\begin{figure}	
	\subfigure[  ]{
		\includegraphics[width=0.225\textwidth, height=0.18\textwidth]{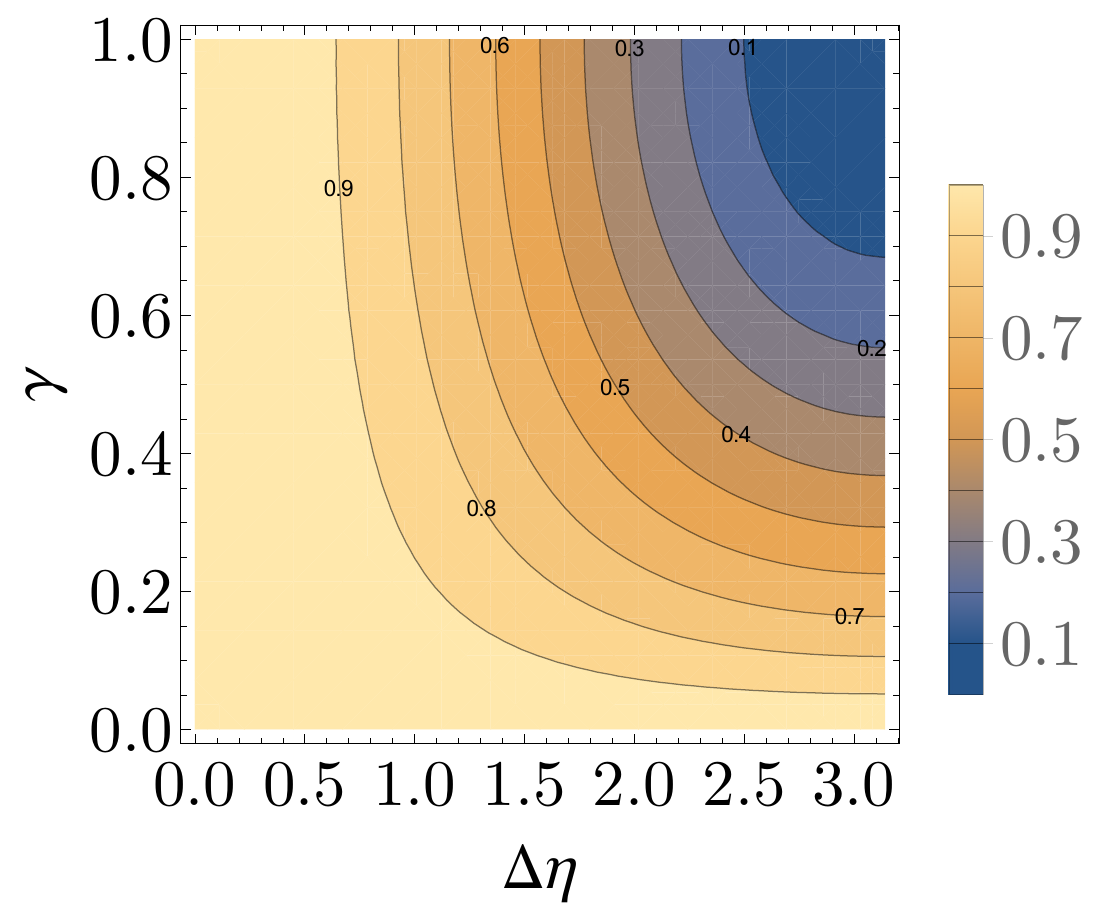}
		\label{fig1}
	}
	\subfigure[ ]{
		\includegraphics[width=0.225\textwidth, height=0.18\textwidth]{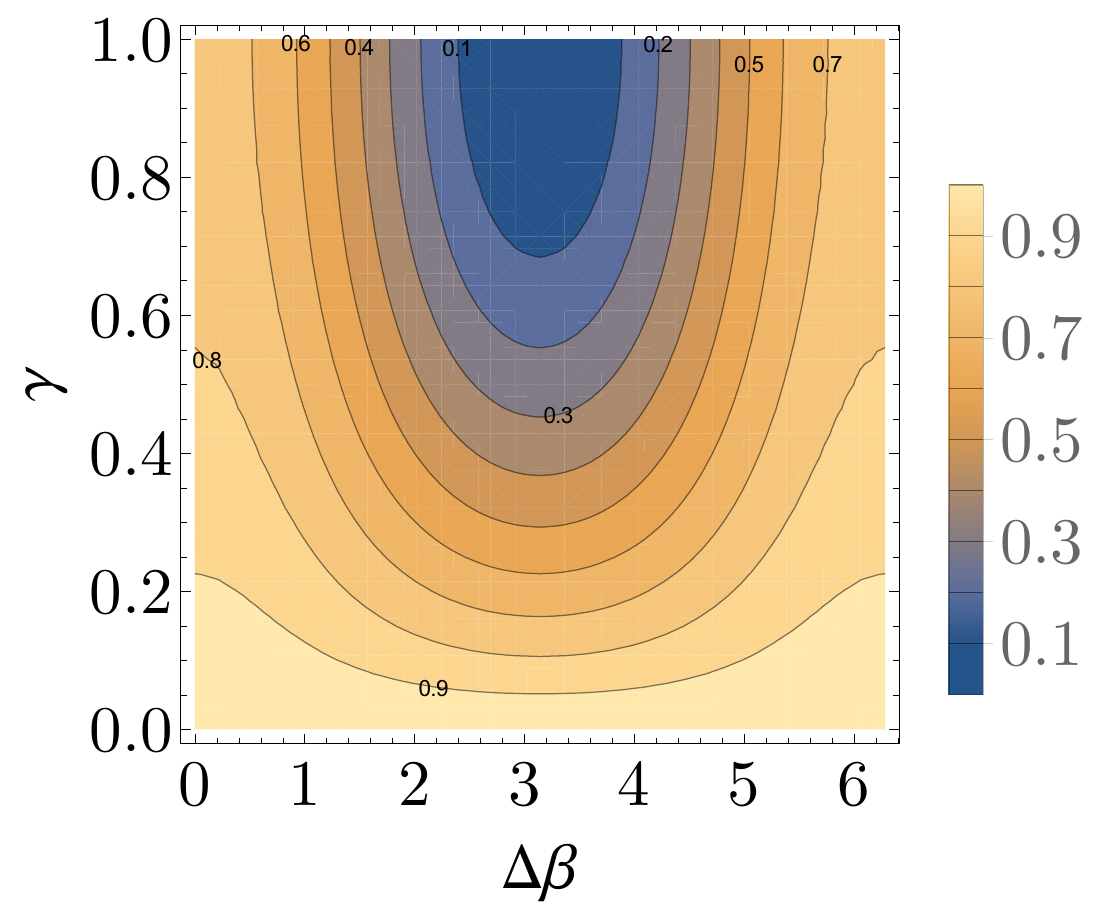}       
		\label{fig2}
	}
\subfigure[]{
	\includegraphics[width=0.225\textwidth, height=0.18\textwidth]{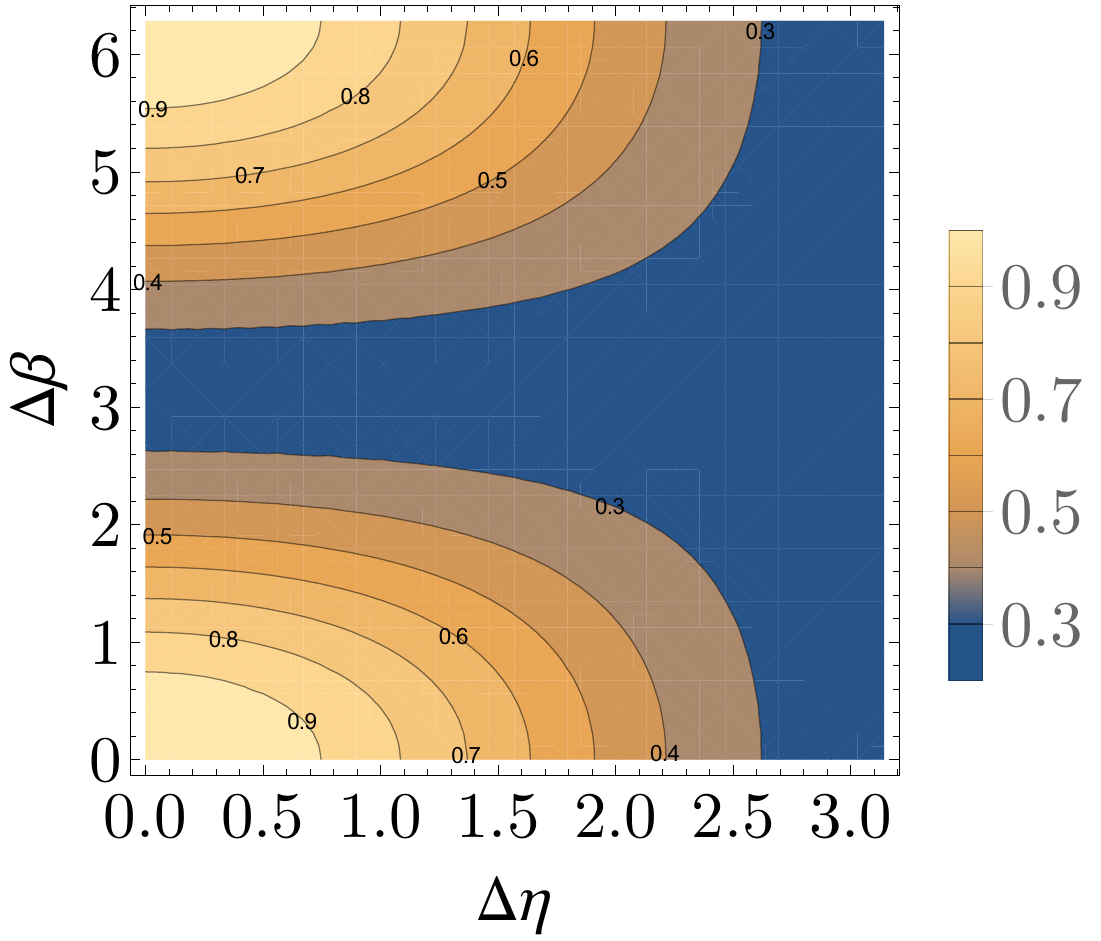}       
	\label{fig5}
}
\caption{(Color online) Contour plots of $\mathcal{F}$  in presence of DC. The maps of the complementarity function is plotted in different planes of parameters by fixing others. In (a): $(\Delta \eta$,   $\gamma$)-plane, with $\Delta \beta=0 $; $\Delta \delta=0$,  (b):  ($\Delta \beta,  \gamma$)-plane, by fixing $\eta_1=0$; $\eta_2=\frac{\pi }{3}$; $\Delta \delta=0$ and (c): ($\Delta \eta, \Delta \beta$)-plane, with  $\Delta \delta=0$; $\gamma =\frac{1}{2}$.  }
\label{f_distribution}
\end{figure}

\emph{Case 1: $\Delta \beta=0 $; $\Delta \delta=0$. } This is the case when both the unitaries have the same phase factors, i.e. \(\beta_1 = \beta_2\), and \(\delta_1 = \delta_2\). Eq. (\ref{dualitydepol}) reduces to
\begin{equation}
\mathcal{F}=1-\frac{\left(2 \gamma- \gamma ^2\right)}{2}  (1-\cos\Delta \eta ),
\label{dc1}
\end{equation}
where $\Delta \eta = \eta_{1} - \eta_{2} $ i.e., $\mathcal{F}$ does not depend individually on $\eta_1$ and $\eta_2$ which is not the case in general. Clearly, \(\mathcal{F} =1\), saturating the inequality, (\ref{func}) when there is no noise i.e., when \(\gamma=0\).   We ignore \(\gamma =2\) as it looses physical interpretation. On the other hand, for \(\Delta \eta = 0 \),  i.e. for
 \(\eta_1 = \eta_2\), the second term  also vanishes and the duality relation saturates for all values of noise parameters, \(\gamma\).  Continuity of \(\Delta \eta\) guarantees that there is a neighborhood where \(\mathcal{F} \approx 1\), $\forall \gamma$. Moreover, we observe that there exists a finite region in the plane  of \(\Delta \eta\), for which \(\mathcal{F} > 0.9\). It indicates that when $\eta_1$ is close to $\eta_2$,  robustness of $\mathcal{F}$ is maximum against DC (see Fig. \ref{fig1}).
 
 Let us enumerate the following observations in this scenario: 
 \begin{enumerate}
 	\item  To keep the value of \(\mathcal{F}\) being fixed to $\mathcal{F}_0$, say, with a specific noise in the detector, one requires a maximum value of \(\Delta \eta\), denoted by \(\Delta \eta_{\max}\) below which \(\mathcal{F}\geq\mathcal{F}_0\),  as  illustrated in the figure by contours. For example, when \(\mathcal{F}\approx0.8 \), and  \(\gamma =0.2\) ,    \(\Delta \eta_{\max} = \) 1.68. 
 	
\item For a fixed value of $\gamma$, let us consider two different values of $\mathcal{F}$, say, ${\mathcal{F}_1}$ and  ${\mathcal{F}_2}$, such that ${\mathcal{F}_1} \geq  {\mathcal{F}_2}$, then \(\Delta \eta_{\max}(\mathcal{F}_1) \leq \Delta \eta_{\max}(\mathcal{F}_2)\). It tells us that corresponding to a large value of $\mathcal{F}$, there exists a small range of $\Delta \eta$, for which $\mathcal{F}$ can be tuned towards saturation of \(\mathcal{F}\). 

\item  For  high values of  \(\gamma\), i.e., in presence of a very  noisy environment,  and when the difference between \(\eta_1\) and \(\eta_2\) is close to \(  \pi\), \(\mathcal{F}\)  goes far from unity, i.e. far from saturation. This is exactly the opposite scenario than the one with $\Delta\eta=0$.
\end{enumerate}



\emph{Case 2: $\eta_1=0$; $\eta_2=\frac{\pi }{3}$; $\Delta \delta=0$ or $\eta_1=0$; $\eta_2=\frac{\pi }{3}$; $\Delta \beta=0$}. 
We choose specific values of  $\eta_i$s and $\Delta\delta=0$  ($\Delta\beta=0$), to study the behaviour of $\mathcal{F}$ with respect to $\Delta \beta$ ($\Delta\delta$) and $\gamma$. Note that  $\eta_1=0$ makes $U_1$ to be diagonal while all the elements in $U_2$ are non-vanishing with  $\eta_2=\frac{\pi}{3}$. Eq. (\ref{dualitydepol}) in this case reduces to
\begin{equation}\label{dc2}
\mathcal{F}=1-
\frac{\left(2 \gamma-\gamma ^2 \right)}{8}  (5-3 \cos \Delta \beta),
\end{equation}
with $\Delta\delta=0$. Similar expression can be found by replacing $\Delta\beta$ by $\Delta\delta$. Since the role of $\Delta \beta$ and $\Delta\delta$ on $\mathcal{F}$ is the same, we will get a similar equation and behaviour when one of them vanishes. We find that the complementarity relation saturates when there is no noise in the system, i.e., $\gamma=0$. However, unlike in the previous case,  there does not exist any value of $\Delta\beta$ for which saturation of $\mathcal{F}$ happens for certain values of \(\gamma\). 
For example, for $\Delta\beta=0$ or $2\pi$, we have $\mathcal{F} \approx 1$, for a certain range of $\gamma$ only. As depicted in Fig. \ref{fig2}, the effect of noise is minimized on $\mathcal{F}$ when $\Delta\beta=0$ or 2$\pi$ and its neighborhood. 

We also find that to keep $\mathcal{F}$ close to a fix value, there exists a finite range of $\Delta\beta$,  for a fixed $\gamma$. As a particular case, when $\mathcal{F}\approx0.8$ and $\gamma=0.15$, $1.83\leq \Delta\beta  \leq 4.45$. Comparing different contours of $\mathcal{F}$, for fixed values of $\gamma$, we observe that with a decrease of $\mathcal{F}$, the range of $\Delta\beta$ also shrinks. For example, for $\mathcal{F}\approx 0.5$ and $\gamma=0.3$, $2.81 \leq \Delta\beta \leq 3.47$. Another sharp contrast between Case 1 and Case 2 is that, for certain $\gamma$ values, tuning $\Delta\beta$ is not enough to obtain a high value of $\mathcal{F}$. Moreover, the complementarity is highly affected by  noise, i.e., $F << 1$, when $\Delta\beta$ is close to $\pi$.

\emph{Case 3: $\Delta \delta=0$; $\gamma =\frac{1}{2}$ or $\Delta \beta=0$; $\gamma =\frac{1}{2}$.} Let us study  \(\mathcal{F}\)  in the \((\Delta \eta, \Delta \beta) \)-plane for a fixed \(\gamma\). We choose \(\Delta \delta =0\). The similar picture for   \(\mathcal{F}\) can be obtained when \(\Delta \beta\) is replaced by \(\Delta \delta\) and \(\Delta \beta\) is fixed to zero. 
In this case, we have 
%
\begin{eqnarray}\label{dc3}
\mathcal{F}&=&\frac{1}{16} \Big(7+3 \cos \Delta \beta\big(1+\cos \Delta \eta\big)  +  3 \cos \Delta \eta \Big). 
\end{eqnarray}
The complementarity  relation saturates or \(\mathcal{F} \approx 1\), if $\Delta \eta=0$ and $\Delta \beta=0$ or  $\Delta \eta =0\) and \(\Delta \beta = 2\pi$ as shown in Fig. \ref{fig5}. 
On the other hand, $\mathcal{F}\approx0$ with $\Delta\beta$ ($\Delta \eta$) is close to $\pi$ irrespective of  \(\Delta \eta \)  ($\Delta\beta$).

\emph{\textbf{Asymmetric Case:}} Let us now  move to the case when the initial quanton state is asymmetric i.e., $p_1\neq p_2$. In this case, the complementary function in (\ref{func}) can be denoted as  $\mathcal{F}_a \equiv {\cal C}_a^2 + {\cal D}_a^2$  where ${\cal C}_a$ and ${\cal D}_a$ are the coherence and path distinguishability with asymmetric quantons. It depends both on $\theta$ and $\phi$. 
For comparison, let us also denote coherence, path distinguishability and complementarity by ${\cal C}_s^2$, ${\cal D}_s^2$, $\mathcal{F}_s^2$ for the symmetric case. 

We find that for fixed unitary operators, path distinguishability increases, i.e., ${\cal D}_s^2\leq {\cal D}_a^2$ while coherence decreases, ${\cal C}_s^2 \geq {\cal C}_a^2$, with the increase in asymmetry. Also, there exists a range of values of \(\theta\)($\phi$) where \({\cal D}_a^2\) remains constant. 
 Interestingly, we observe that, when $\mathcal{F}_a >\mathcal{F}_s$, there exists a finite range of $\theta (\phi)$ where  \({\cal D}_a^2\) is independent of \(\theta\)($\phi$)   while for the remaining values of  $\theta (\phi)$, $\mathcal{F}_a =\mathcal{F}_s$ (see Fig. \ref{fig6} and \ref{fig7}). In the region where $\mathcal{F}_a >\mathcal{F}_s$, $\mathcal{F}_a$ has a bump with respect to $\theta$($\phi$).
 This bump appears because $\mathcal{D}_a^2$ becomes constant for a certain range of $\theta$($\phi$), which is a consequence of the fact that distinguishability is lower bounded by the predictability $\mathcal{P}$. At $p_1=\frac{1}{8}$, distinguishability reads in terms of $\theta (\phi)$ as
\begin{equation*}
\mathcal{D}_a^2(\theta) =\begin{cases}
0.5625, \hspace{2cm} \text{for  $ 0.12 \leq \theta \leq 1.37$},\\
0.003\big(204.1-4.1\cos2\theta-51.73\sin2\theta\big), \\ \hspace{3cm} \text{for $\theta \in  [0,0.12)\cup (1.37,\pi]$}.
\end{cases}
\end{equation*}

\begin{equation*}
\mathcal{D}_a^2(\phi) =\begin{cases}
0.5625, \hspace{2cm} \text{for  $\phi \in  [0,0.98)\cup (5.3,2\pi]$},\\
0.006\big(107-28\cos\phi-7\cos2\phi\big), \\ \hspace{3cm} \text{for $ 0.98 \leq \phi \leq 5.3$}.
\end{cases}
\end{equation*}
The result shows that manipulation in the detector state with asymmetric quanton can push the complementarity towards saturation. The physical origin of the advantage obtained with asymmetric quanton states is as follows:  With increase in asymmetry of the states, both distinguishability and predictability increase while the coherence decreases. It so happens that, the loss in coherence for certain values of the initial state parameters is overcompensated by the increase in the distinguishability. For these initial states, there is sharp rise in the value of predictability, $\mathcal{P}$, and hence the distinguishability, $\mathcal{D}$, which leads to increment in the complementarity relation compared to that in the symmetric case.

\subsection{Amplitude damping channel}
\label{subsec_ampd}

\begin{figure}[H]
	\subfigure[]{
		\includegraphics[width=0.225\textwidth, height=0.17\textwidth]{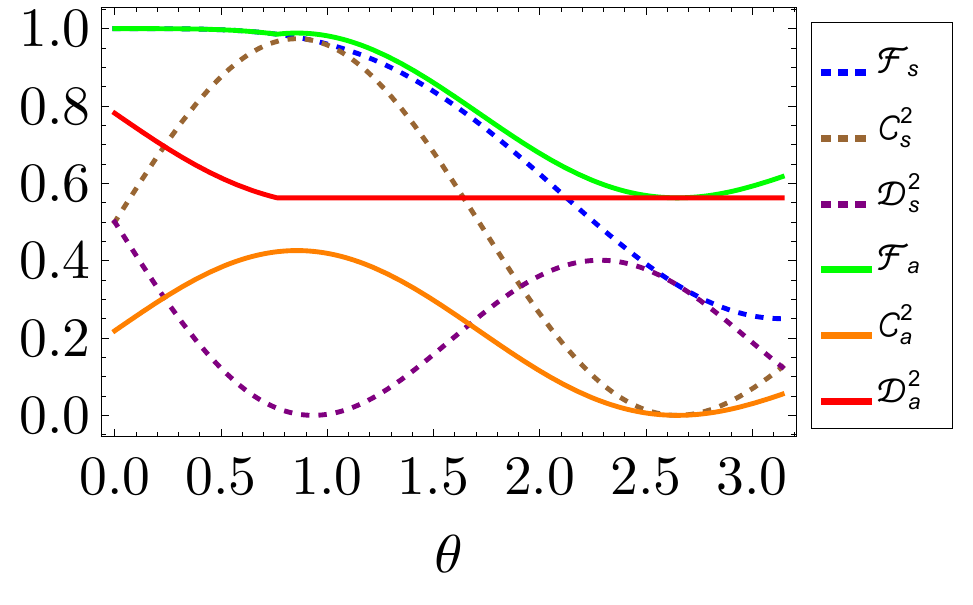}
		\label{adc6}
	}
	\subfigure[]{
		\includegraphics[width=0.225\textwidth, height=0.17\textwidth]{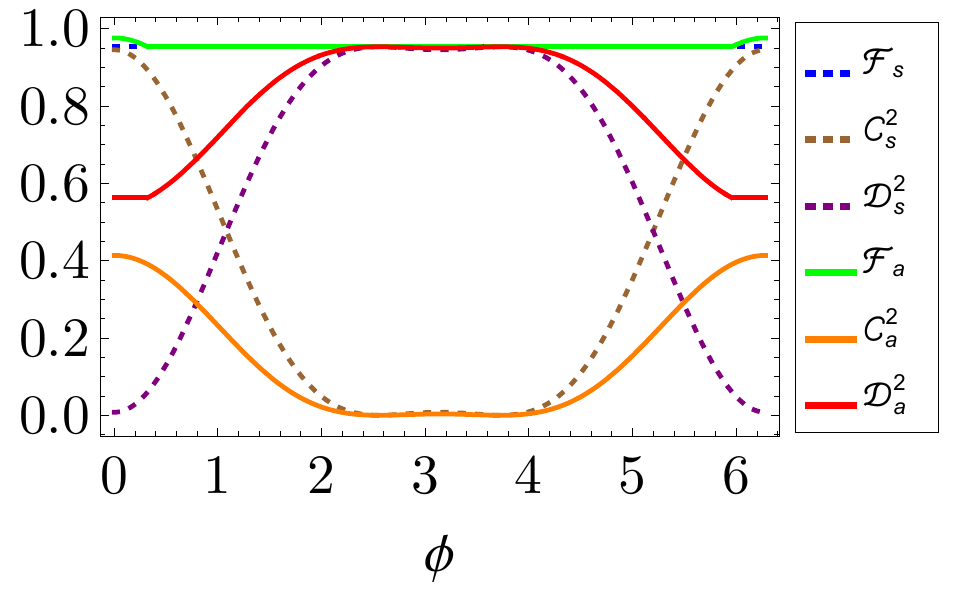}       
		\label{adc7}
	}
	\caption{(Color online) Like in Fig. \ref{f_distr}, complementarity function is plotted for ADC. Other parameters are fixed to
$\text{$\eta_1$}=\pi ;\text{$\eta_2$}=\frac{\pi }{2};\text{$ \Delta \beta $}=2 \pi ;\text{$\Delta \delta $}=\pi; \gamma =\frac{1}{4}$ in both figures, with $\phi =0$ in (a) and $\theta =\frac{\pi }{3}$ in (b). As in the case of depolarizing noise, asymmetric quanton (with $p_1=\frac{1}{8}$) outperforms for some finite range of \(\theta\) values over the symmetric one.  }
	\label{f_dis}
\end{figure}

Let us now investigate the consequence on complementarity relation when amplitude damping channel (see \ref{subsec:ampdampK}) acts on the detectors. Let us again take the initial quanton state of the form $\ket{\psi_{in}}=\frac{1}{\sqrt{2}}(\ket{\psi_{1}}+\ket{\psi_{2}})$. In this case, the  duality relation in (\ref{func}) takes the form as
\begin{widetext}
	\begin{align}
	\mathcal{F}=1+\frac{9}{8}\gamma(\gamma-1)+\frac{\gamma(\gamma-1)}{8}\Big(3\cos 2\theta-12\cos \theta-8\big(\cos \eta_1 \cos \eta_2  -(\cos \eta_1+\cos\eta_2)\sin\Delta \beta\sin\Delta \delta \nonumber \\
	+\cos\Delta \delta\sin\eta_1 \sin\eta_2+\cos\Delta \beta\big[\cos\Delta \delta+\cos\Delta \delta\cos\eta_1\cos\eta_2+\sin\eta_1\sin\eta_2\big]\sin^4\frac{\theta}{2}\big)\Big)\leq 1.
	\end{align}
\end{widetext}
Like the DC, we first note that $\mathcal{F}$ depends on $\Delta \beta$, $\Delta \delta$, $\eta_1$, $\eta_2$ and $\gamma$. Moreover, $\mathcal{F}$ is a function of the detector state parameter, $\theta$. Unlike the depolarizing noise, we find that $\mathcal{F} \approx 1$ when   
\(\theta\) is close to zero for other fixed parameters and for moderate amount of noise. Such observation holds both for the initial quantum states with \(p_1 = p_2\) and with \(p_1 \neq p_2\)  (see Fig. \ref{adc6}). We again see that the asymmetric quanton states can perform better than that of the symmetric ones with respect to the saturation for certain choices of \(\theta\). The state having \(p_1 \neq p_2\)  shows advantage again due to the behaviour of path-distinguishability  with the variation of \(\theta\) as it is clearly visible from Fig. \ref{adc6}. On the other hand, \(\mathcal{F}\) with \(p_1 = p_2\) remains constant with \(\phi\) in \(|d_0\rangle\) while it shows nonmonotonicity with \(\phi\) for the case when \(p_1\) is different from \(p_2\) for the initial state (see Fig. \ref{adc7}). 

To illustrate the behaviour of \(\mathcal{F}\) with \(\theta\) and \(\gamma\), we now perform the following analysis for the initial quanton with \(p_1 = p_2\). 

\begin{figure}[H]	
	\subfigure[]{
		\includegraphics[width=0.225\textwidth, height=0.18\textwidth]{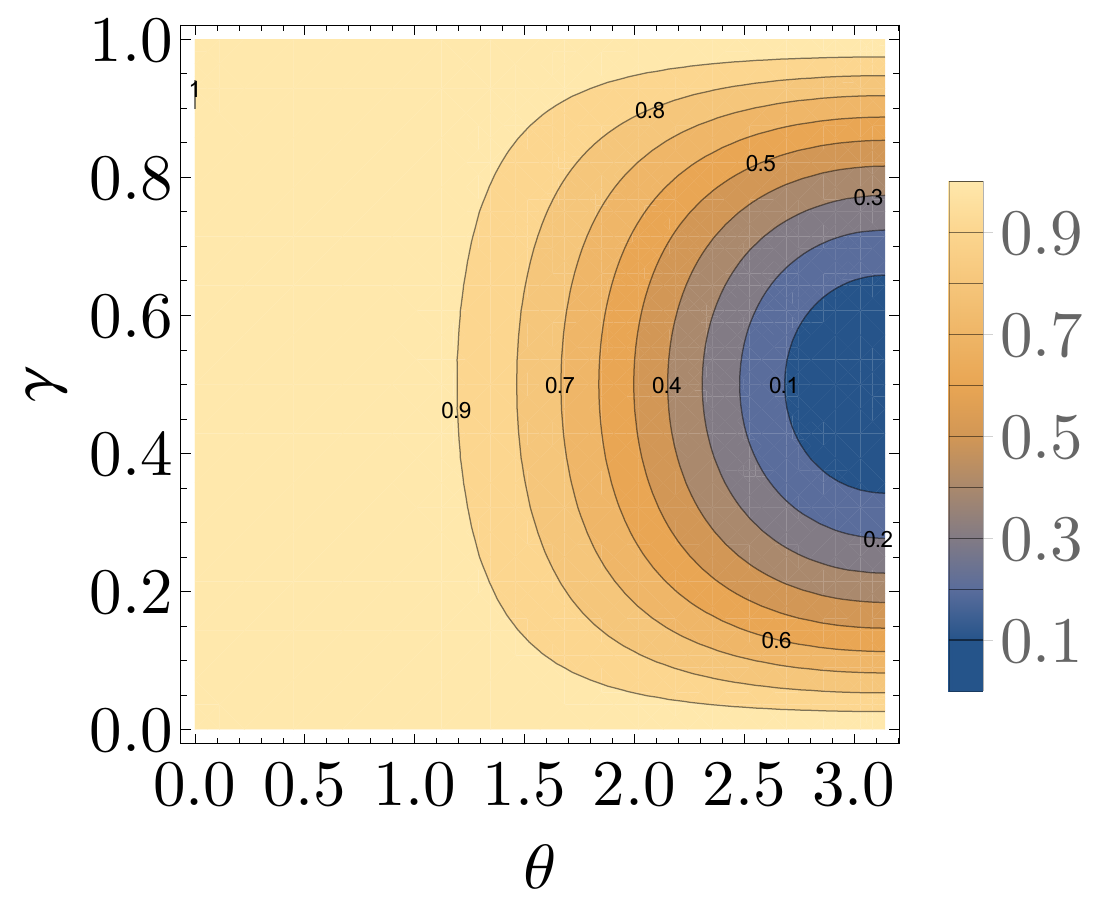}
		\label{adc1}
	}
	\subfigure[]{
		\includegraphics[width=0.225\textwidth, height=0.18\textwidth]{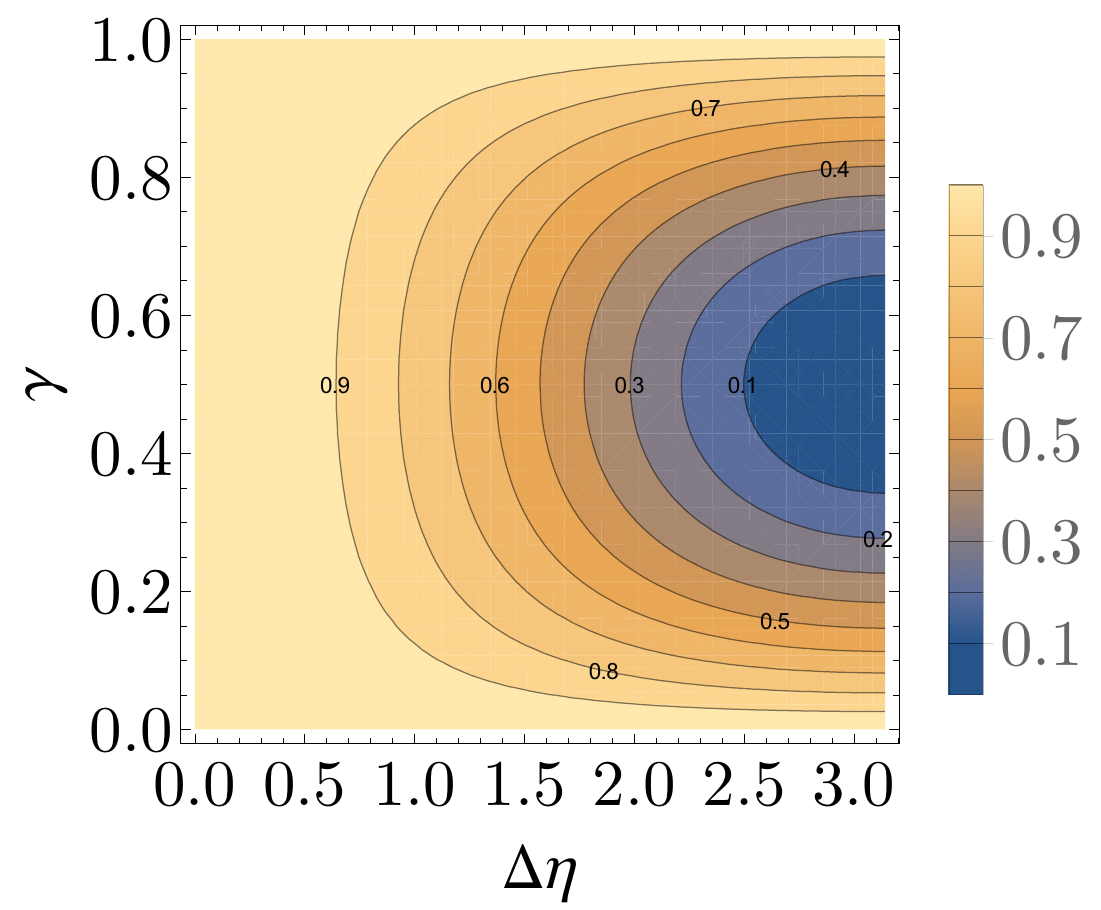}       
		\label{adc2}
	}
	\subfigure[]{
		\includegraphics[width=0.225\textwidth, height=0.18\textwidth]{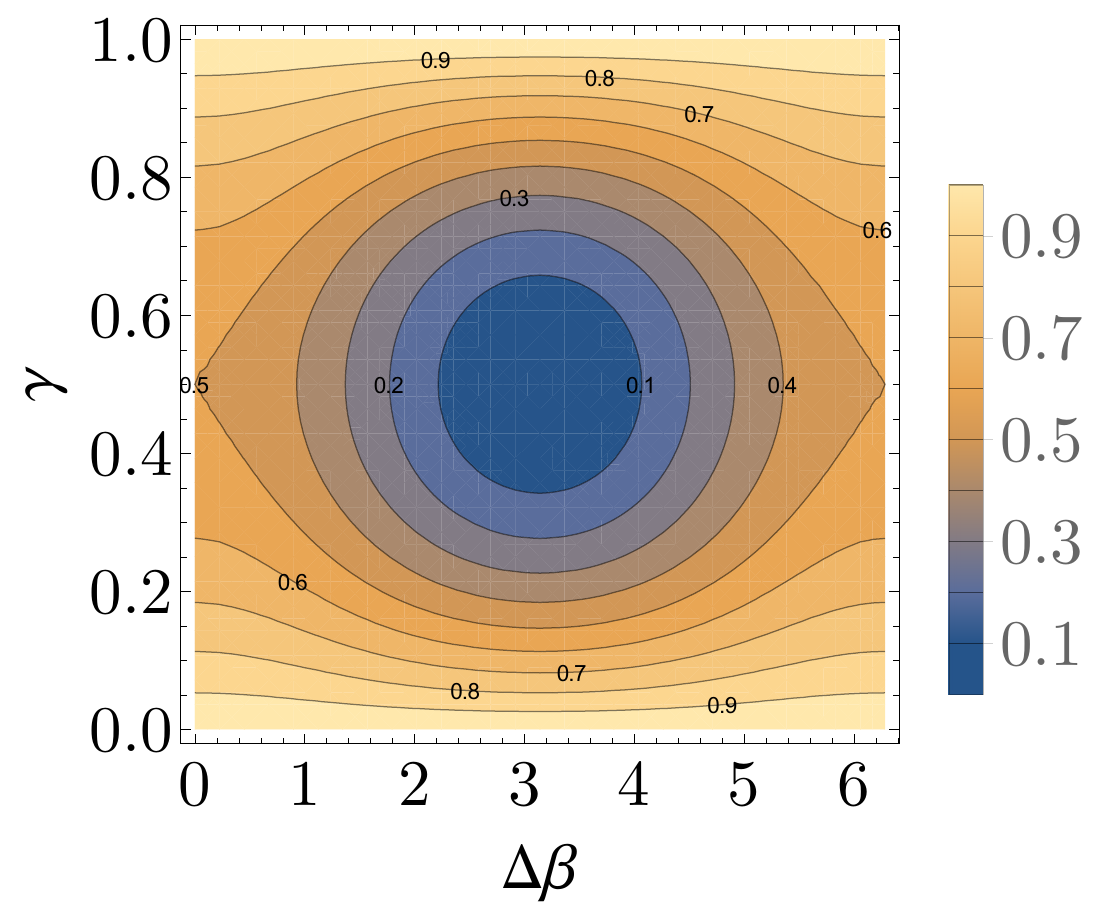}       
		\label{adc3}
	}
    \subfigure[]{
    	\includegraphics[width=0.225\textwidth, height=0.18\textwidth ]{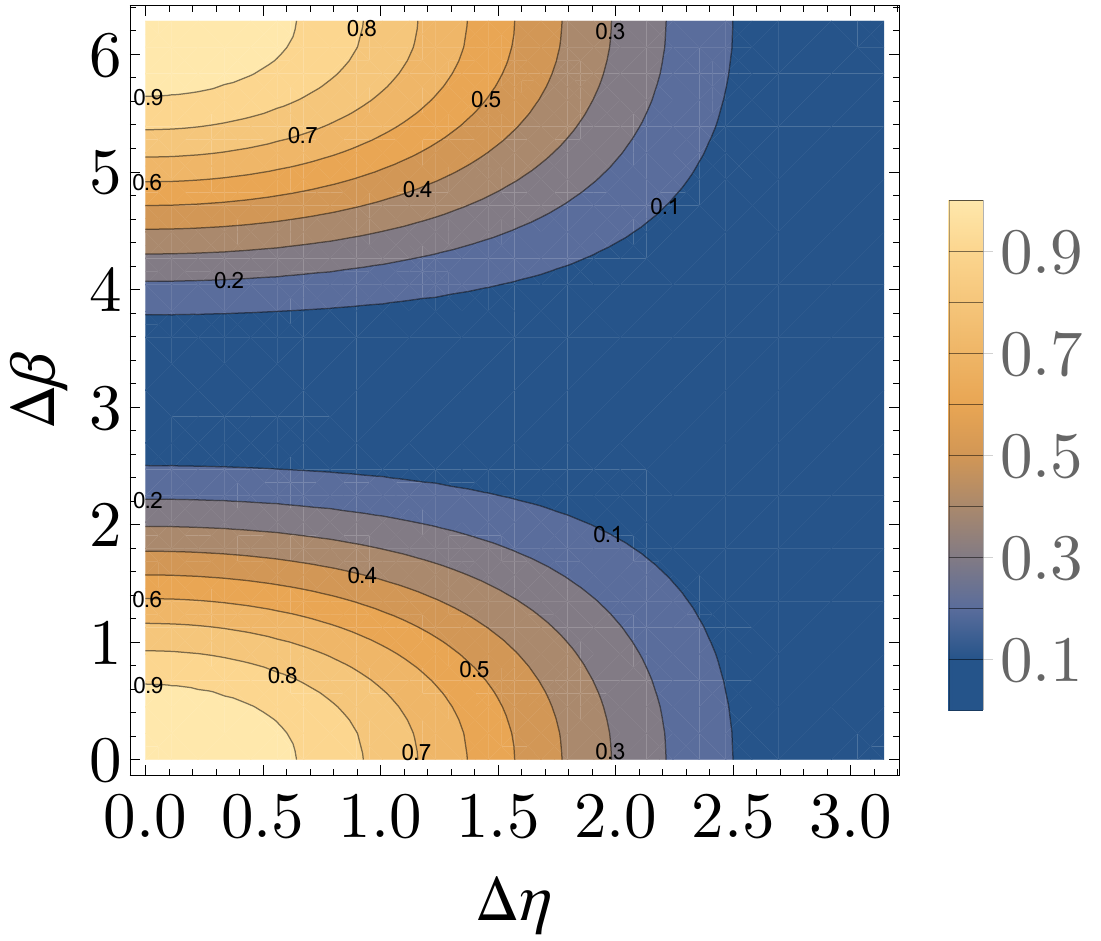}       
    	\label{adc4}
    }
	\caption{(Color online) Contour plots of $\mathcal{F}$ in the case of ADC.  The horizontal and the vertical axes respectively are in (a) $\gamma$ and $\theta$, with $\Delta \beta=0$; $\Delta \delta=0$; $\eta_1 $=0; $\eta_2 =\pi$, (b) $\gamma$ and $\Delta \eta$, having $\Delta \beta=0$; $\Delta \delta=0$; $\theta =\pi$, (c) $\gamma$ and $\Delta \beta$, $\Delta \delta=2\pi$; by fixing $\eta_1 $=0; $\eta_2$=$\frac{\pi }{2}$; $\theta =\pi$ and (d) $\Delta \eta$ and $\Delta \beta$, with  $\Delta \delta=0$; $\gamma =\frac{1}{2}$ and $\theta=\pi$.}
	\label{f_dist}
\end{figure}

\emph{Case 1: $\Delta \beta=0$; $\Delta \delta=0$; $\eta_1 $=0; $\eta_2 =\pi$.} These choices of parameters fix  both the unitaries and the expression for $\mathcal{F}$,  which reduces to
\begin{equation}\label{ac1}
\mathcal{F} =\frac{1}{2} \Big(2 -(\gamma-\gamma^2 ) (3-4 \cos \theta +\cos 2 \theta )\Big). 
\end{equation}
 Note that $\mathcal{F}=1$, when  $\gamma=0$ or 1 \(\forall \theta\)  and  $\theta=0$ $\forall$ $\gamma$.  It implies that for $\theta=0$, i.e., when $\ket{d_0}=\ket{\psi_{1}}$, the complementarity can always be saturated for any value of the noise parameter $\gamma$, which indicates that the complementarity relation is maximally robust against noise when  $d_0 = |\psi_1\rangle$.  However, when \(|\psi_{in}\rangle =\frac{1}{\sqrt{2}} (|\psi_1\rangle + |\psi_2\rangle) \) and \(|d_0 \rangle = | \psi_2\rangle\), the ADC has maximal destructive effect on \(\mathcal{F}\). Such observation can be justified from the definition of ADC, given in Eqs. (\ref{12}).   
 We observe that $\mathcal{F}\geq0.9$ when $0\leq\theta\leq\pi/2$ in presence of any strength of the noise. More specifically, for fixed  $\gamma=0.075$, we  obtain $\mathcal{F}\geq0.9$,  for  $\theta \leq 1.77$. In the presence of fixed noise parameter, such an upper bound on \(\theta\) can always be found which gives a certain fixed value of \(\mathcal{F}\). We depict this feature by using contours in \( (\theta, \gamma) \)-plane. 
 Further, the value of $\mathcal{F}$ goes close to zero or the complementarity remains far from saturation, if $\gamma=\frac{1}{2}$ and $\theta={\pi}$ (see Fig. \ref{adc1}). 


\emph{Case 2:  $\Delta \beta=0$; $\Delta \delta=0$; $\theta =\pi$}. Such a choice makes the detector state to be \(|d_0 \rangle = |\psi_2\rangle\) while the initial state is the symmetric one. For some unitary operators, our analysis in Case 1 shows that such a choice of \(\theta\) pushes away \(\mathcal{F}\) from unity. The question remains in this case whether we can tune the unitaries in such a way that the trigonometric function \(\eta_i\)s change and \(\mathcal{F}\) goes towards saturation. 
With these choices of parameters, we obtain
\begin{equation}\label{ac2}
\mathcal{F}  = 1-2  (\gamma-\gamma^2 )  \big(1-\cos \Delta \eta\big).
\end{equation}
From the above expression, we see that $\mathcal{F}$ saturates, for $\gamma=0$ or $1$ \(\forall \Delta \eta\)  and  $\Delta\eta=0$ $\forall \gamma$. Moreover, there is a certain range of $\Delta\eta$ close to zero, in which $\mathcal{F}\approx 1$ for all values of $\gamma \leq 1/2$ (see Fig. \ref{adc2}). On the other hand, for small values of \(\gamma\), any \(\Delta \eta\) values leads to the saturation of \(\mathcal{F}\) as depicted in Fig. \ref{adc2}.  



\emph{Case 3: $\Delta \delta=2\pi$; $\eta_1 $=0; $\eta_2$=$\frac{\pi }{2}$; $\theta =\pi$ or $\Delta \beta=2\pi$; $\eta_1 $=0; $\eta_2$=$\frac{\pi }{2}$; $\theta =\pi$.} To see the effect of phase $\Delta \beta$ or $\Delta \delta$ in the unitary on $\mathcal{F}$, we choose some specific values of  parameters. With $\Delta\delta=2\pi$,
\begin{equation}\label{ac3}
\mathcal{F} =  1-(\gamma-\gamma^2 ) \big(3-\cos \Delta \beta\big).
\end{equation}
From the above equation, we find that  $\mathcal{F}=1$ for $\gamma=0$ or 1 and \(\mathcal{F} =0\) for \(\gamma =0.5\) and \(\Delta \beta = \pi\). As we infer from Fig. \ref{adc3}, \(\mathcal{F}\) can only go to saturation for all values of \(\Delta \beta\) when \( \gamma\) is very small, i.e., the  noise is almost negligible. Moreover, we see that the pattern of \(\mathcal{F}\) with \( \Delta \beta\) for \(1/2 \leq \gamma \leq 1\) is the mirror reflection of the one obtained with  \(0 \leq \gamma \leq 1/2\). As it is clear from Eq. (\ref{ac3}), with increasing \(\gamma\), \(\mathcal{F} \) decreases. However for fixed \(\gamma\), to obtain  \(\mathcal{F} < 0.5\), \(\Delta \beta \)  is in the neighborhood to \(\pi\) while to obtain  \(\mathcal{F}  \approx 0.5\), \(\Delta \beta\) should be taken towards zero or \(2 \pi\). 

%

If we fix \(\theta =\pi\), \(\Delta \delta =0\) and \(\gamma=1/2\), \(\mathcal{F}\) behaves quite similar to the depolarizing noise in the \((\Delta \eta, \Delta \beta)\)-plane. The functional form in this case is given by 
\begin{equation}
\mathcal{F} = \cos^2 \frac{\Delta \beta}{2}  \cos^2 \frac{\Delta \eta}{2}.  
\end{equation}
Clearly it vanishes when either \(  \Delta \beta = \pi \) or \( \Delta \eta = \pi\) while it saturates when one of the phase is zero. The plot for this scenario is depicted in Fig. \ref{adc4}. From the analysis of \(\theta\) as also seen from Fig. \ref{adc1}, one expects  that \(\mathcal{F}\) has very low values when \(\theta \) is close to \(\pi\) in presence of high noise. The Fig. \ref{adc4} shows that even in this scenario, one can choose \(\Delta \eta\) and \(\Delta \beta\) so that \(\mathcal{F}\) can reach to a very high value, close to saturation.  

On the other hand, if we consider \(\theta =\pi/2\) and choose the same values of  \(\Delta \delta =0\) and \(\gamma=1/2\), then we have 
\begin{equation}\label{ac5}
\mathcal{F} =   \frac{1}{16}(13+\cos\Delta\eta+\cos\Delta\beta(1+\cos\Delta\eta)), 
\end{equation}
which has the minimum value \(0.75\) when \(\Delta \eta \) or \(\Delta \beta\) are chosen to be \(\pi\). Such choice of \(\theta\) turns out to be extremely good to suppress the effects of noise. Comparing Eq. (\ref{dc3}) with Eq. (\ref{ac5}), we find that with the proper choice of \(\theta\),  ADC can produce higher value in \(\mathcal{F}\)  compared to  the DC with respect to \(\Delta \eta \) and \(\Delta \beta\). 


\subsection{Phase damping channel}
\label{subsec_phasedamp}

Taking the symmetric quanton state, we evaluate $\mathcal{F}$ for phase damping channel and find that like ADC, it depends on all the parameters involved in the process except for the phase of $\ket{d_0}$, $\phi$. In this scenario, we get
\begin{widetext}
	\begin{align}
	\mathcal{F}=&1+\frac{3}{8}\gamma(\gamma-2)-\frac{\gamma(\gamma-2)}{8}\Big(3\cos 2\theta+\cos\Delta\beta\cos\Delta\delta(1-\cos 2\theta-2\sin^2\theta\big((\cos\eta_1+\cos\eta_2)\sin\Delta\beta\sin\Delta\delta\nonumber \\ 
	&-\cos\eta_1\cos\eta_2[1+\cos\Delta\beta \cos\Delta\delta] -\sin\eta_1\sin\eta_2[\cos\Delta\beta+\cos\Delta \delta])\big)\Big).
	\end{align}
\end{widetext}
For \(|\psi_{in}\rangle = \sqrt{p_1} |\psi_1\rangle +\sqrt{p_2} |\psi_2\rangle  \) with $p_1\neq p_2$, $\mathcal{F}$ again shows non-monotonic behaviour with $\theta$ and $\phi$  and it also possesses higher value than the symmetric quantum states for a certain range of $\theta$ and $\phi$.

%


\begin{figure}[H]	
	\subfigure[]{
		\includegraphics[width=0.225\textwidth, height=0.17\textwidth]{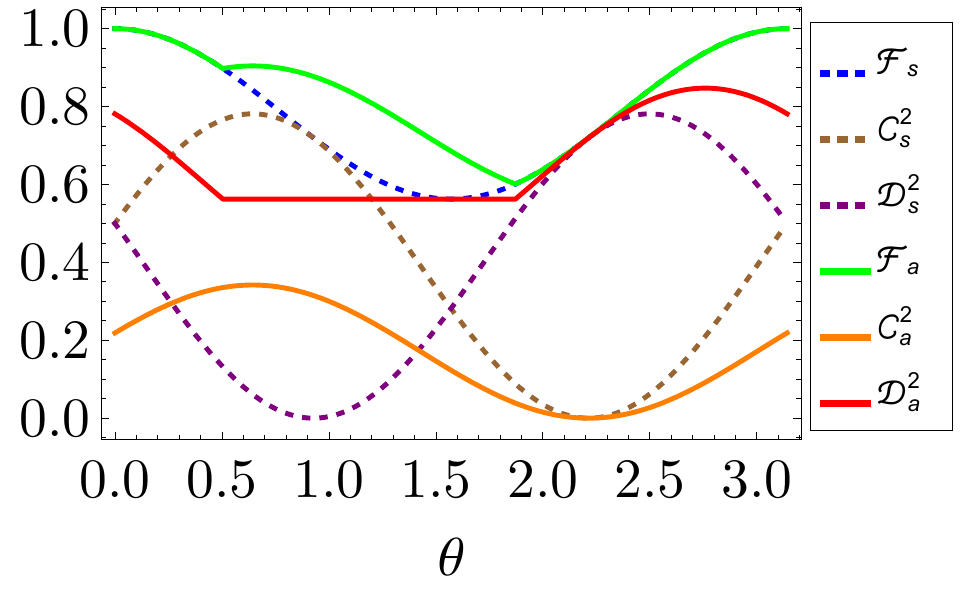}
		\label{dp6}
	}
	\subfigure[]{
		\includegraphics[width=0.225\textwidth, height=0.17\textwidth]{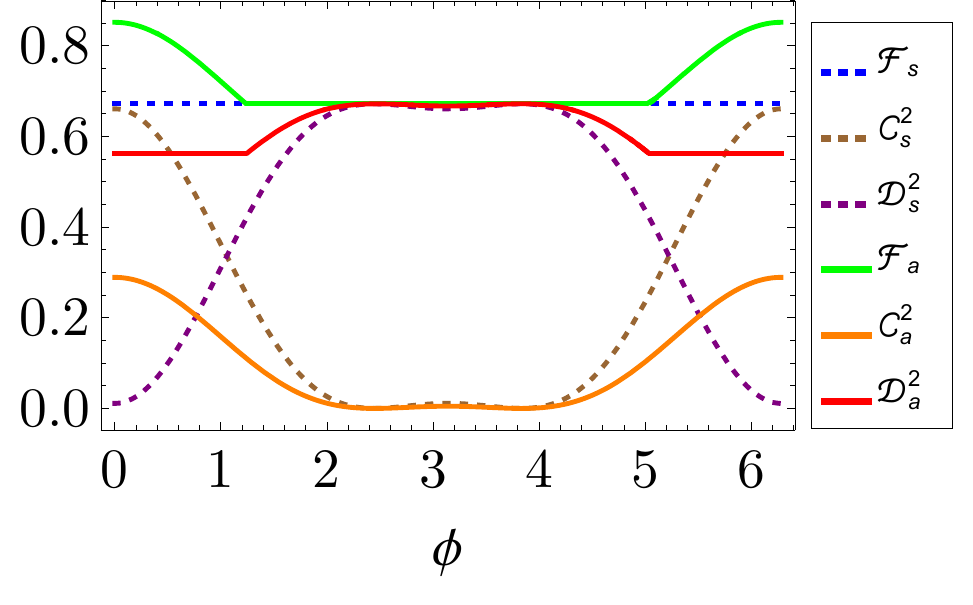}       
		\label{dp7}
	}
	\caption{(Color online) Effects of PDC on \(\mathcal{F}\) against  \(\theta\) (in (a)) and \(\phi\) (in (b)). Quantities and all the other parameters  are same as in Fig. \ref{f_dis}. }
	\label{f_distrer}
\end{figure}Therefore we have,

\emph{Case 1: $\Delta \beta=0$; $\Delta \delta=0$; $\text{$\eta_1$}=0$; $\text{$\eta_2$}=\pi$.}
For these choices, the complementarity relations reads as
\begin{equation}\label{df1}
\mathcal{F}  =   \frac{1}{2} \Big(2-(2\gamma-\gamma^2 )   \big(1-\cos 2 \theta \big)\Big).
\end{equation}
Note that, $\mathcal{F}$ has different functional form than the one in Eq. (\ref{ac1}). In particular, the linear term in Eq. (\ref{ac1}) is missing in this case.

Again, $\mathcal{F}=1$, when $\gamma=0$ or 2 and when $\theta=0$ or $2\pi$. Therefore, the complementarity saturates, which is independent of the noise when we fix the detector state with $\theta=0$ or $\pi$ $\forall$ $\gamma$. Continuity with respect to $\theta$ assures that when  $\theta$  is close to 0 or $2\pi$, $\mathcal{F}\approx 1$. Further, when $\theta=\frac{\pi}{2}$ the effect of noise is most prominent. 
Similar to the ADC, for a fixed value of $\gamma$, there is a range of $\theta$ in which $\mathcal{F}$ can be kept approximately constant. For example $\mathcal{F}\approx 0.9$ with $\gamma=0.075$, when $0.81\leq\theta\leq2.33$. The range shrinks drastically when $\gamma\geq0.5$.
%

\emph{Case 2: $ \Delta \beta=0; \Delta \delta=0;\theta =\frac{\pi }{2}$.}
In this situation, we have 
\begin{equation}\label{df2}
\mathcal{F}=1-\frac{\left(2 \gamma- \gamma ^2\right)}{2}  (1-\cos\Delta \eta ),
\end{equation}
which is exactly similar to the one that we got in the case of DC (see Eq. (\ref{dc1})). However, there is a difference between DC and PDC --  Eq. (\ref{dc1}) is true for all detector states, i.e., for all values of $\theta$ while in case of PDC, the detector state is chosen to be the same as the initial  quantum state in Eq. (\ref{df2}). For example, when $\theta =\frac{\pi}{3}$, the complementarity changes to
\begin{equation}
\mathcal{F}=1-\frac{3\left(2 \gamma- \gamma ^2\right)}{8}  (1-\cos\Delta \eta ).
\end{equation}
Note that for $\gamma=1$ and $\Delta \eta =0$, $\mathcal{F}=\frac{5}{8}$ while for the same choice of values, Eq. (\ref{df2}) gives the value of $\mathcal{F}$ to be $\frac{1}{2}$.
 
\begin{figure}[h]	
	\subfigure[]{
		\includegraphics[width=0.225\textwidth, height=0.18\textwidth]{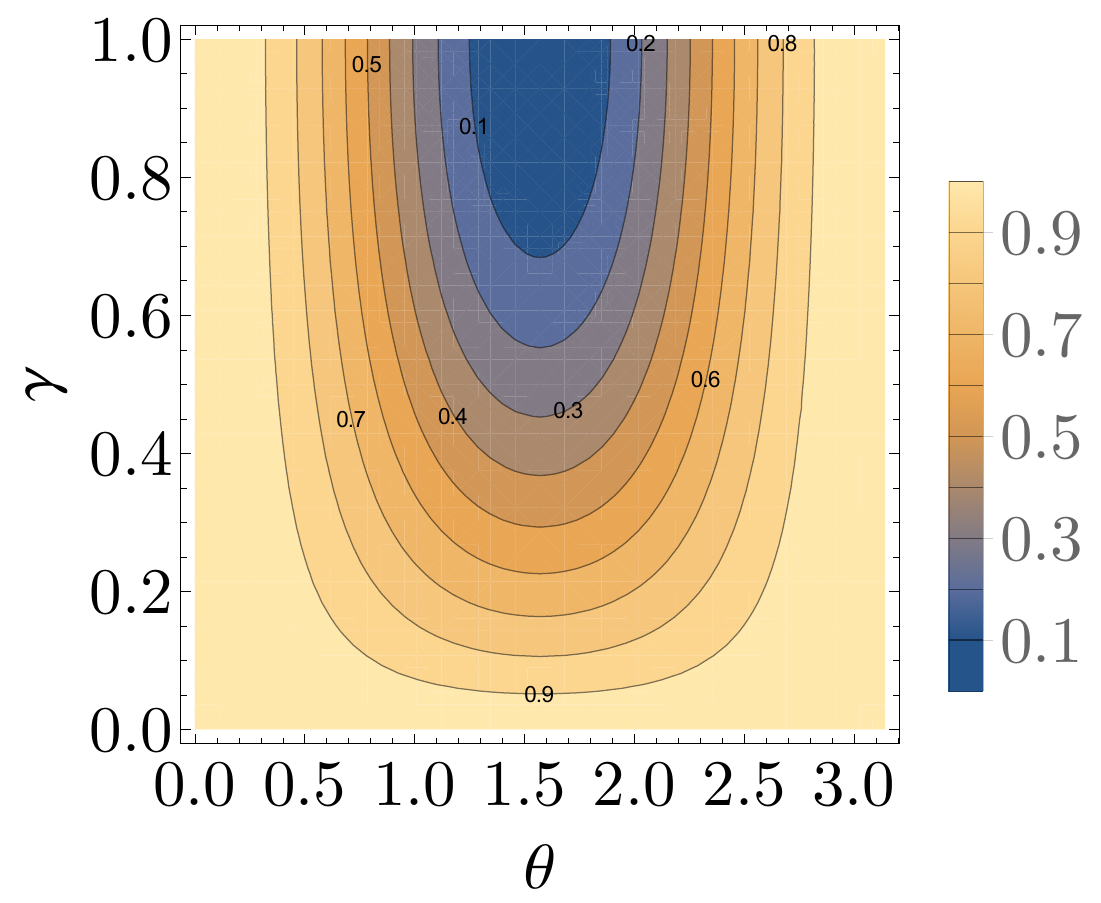}
		\label{dp1}
	}
	\subfigure[]{
		\includegraphics[width=0.225\textwidth, height=0.18\textwidth]{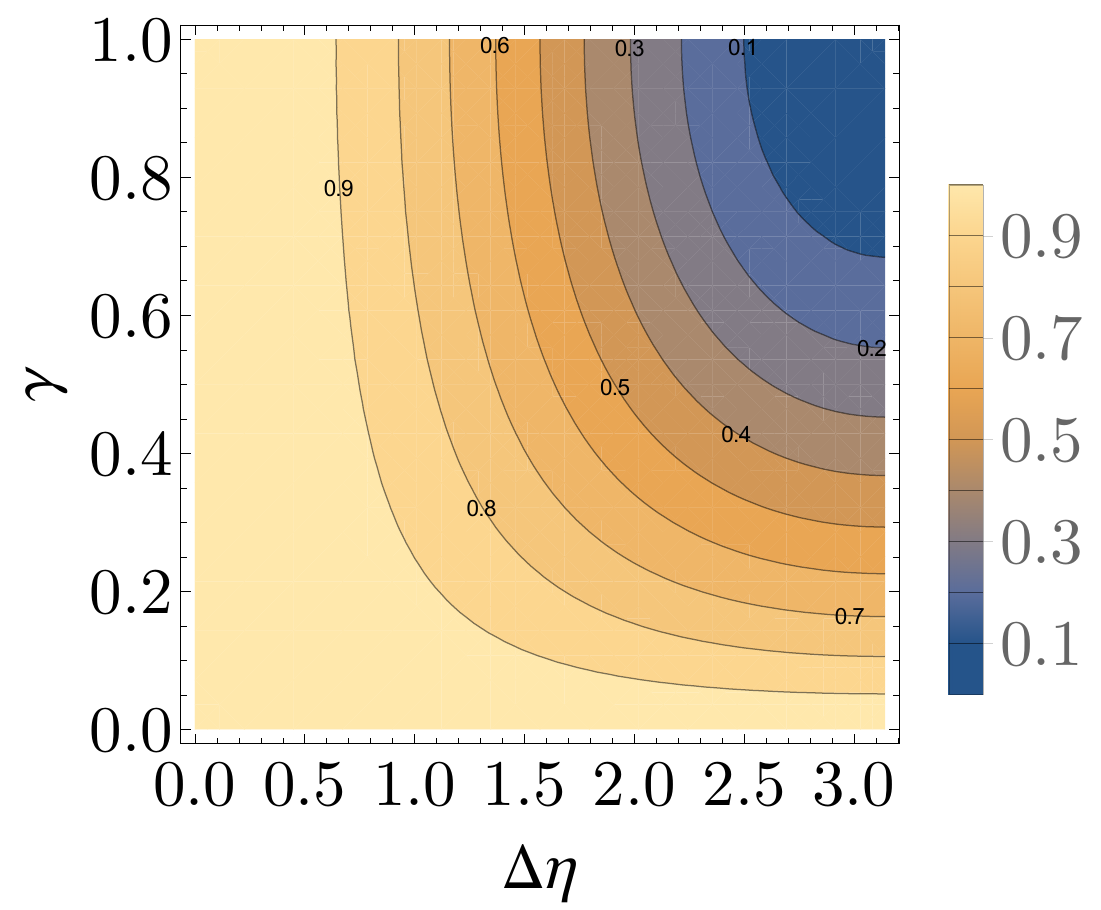}       
		\label{dp2}
	}
	\subfigure[]{
		\includegraphics[width=0.225\textwidth, height=0.18\textwidth]{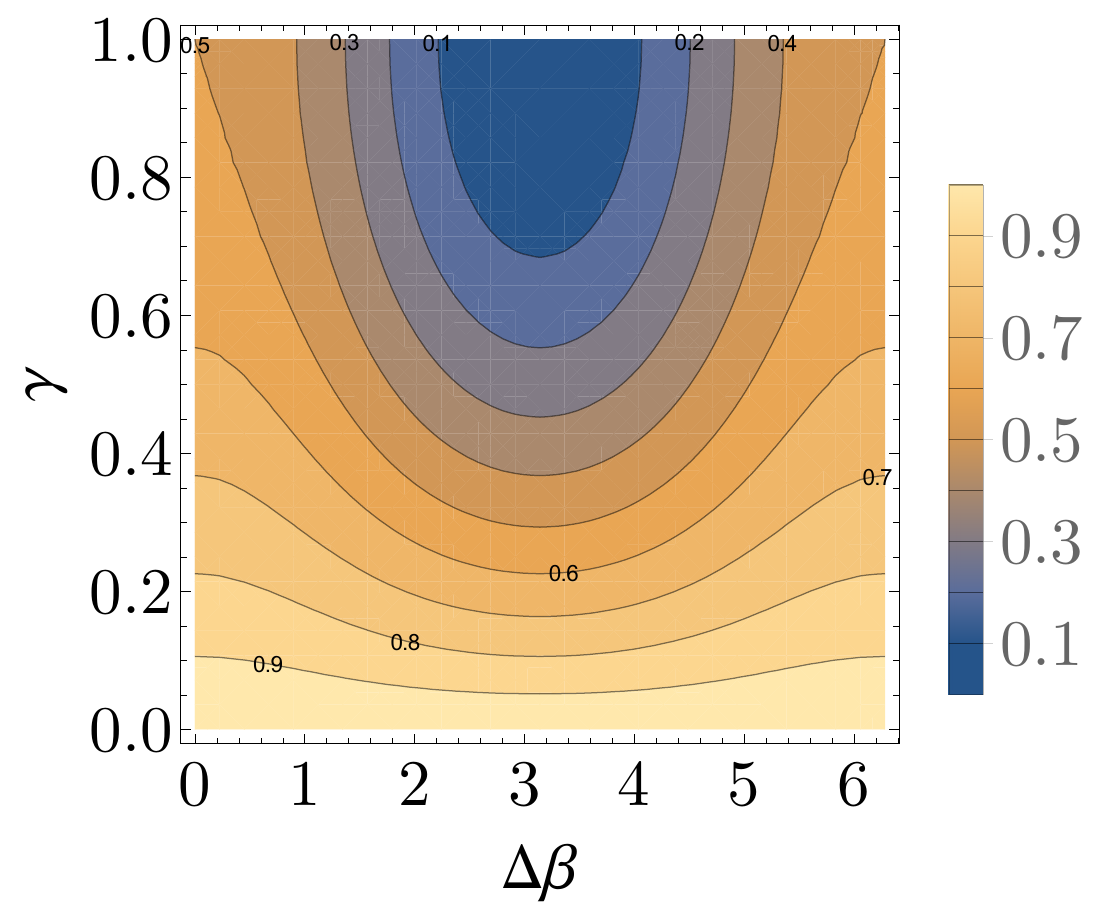}       
		\label{dp3}
	}
\subfigure[]{
	\includegraphics[width=0.225\textwidth, height=0.18\textwidth]{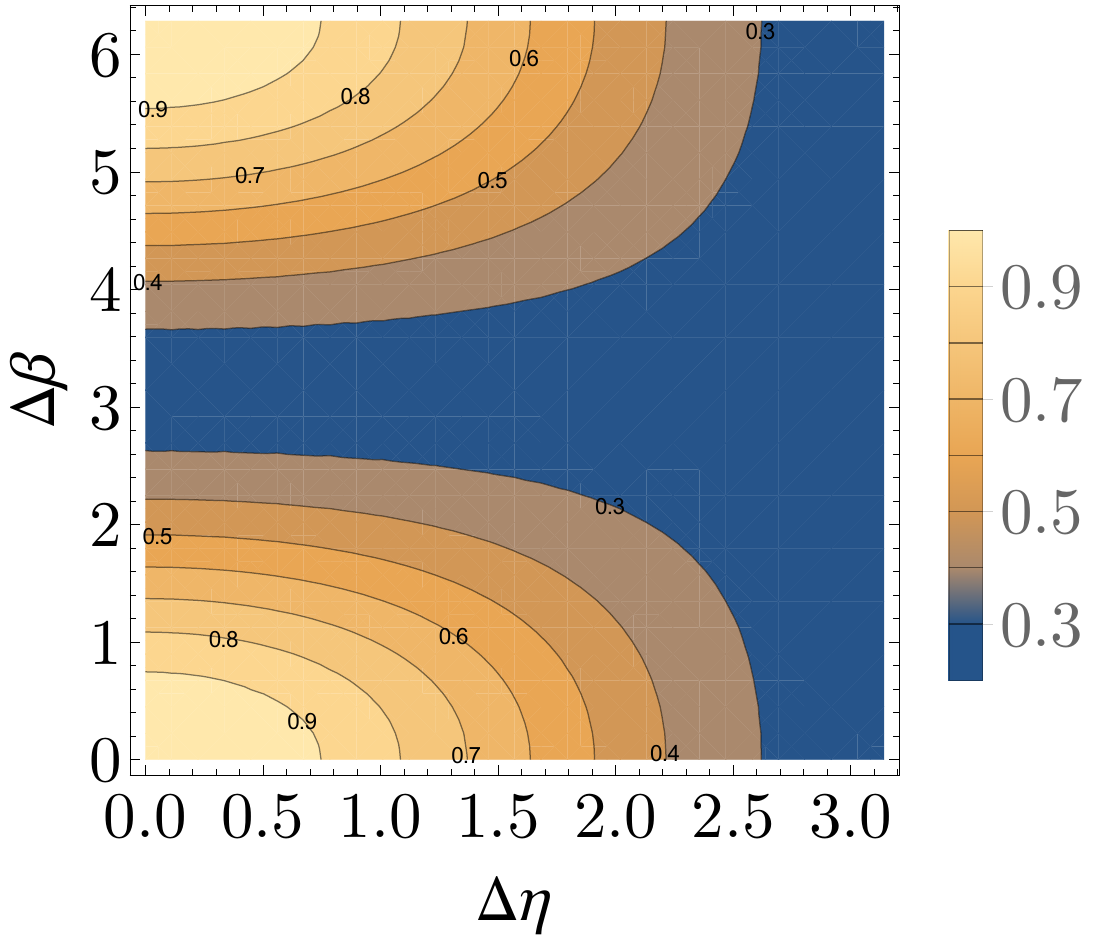}       
	\label{dp4}
}
	\caption{(Color online)  Contour plots of $\mathcal{F}$  when the detector state is affected in PDC. The quantities plotted and the parameters fixed are same as in Fig. \ref{f_dist} except in (b), (c) and (d), \(\theta\) is set to \(\pi/2\). } 
		\label{f_di}
\end{figure}

\emph{Case 3: $\text{$\Delta \delta$}=2 \pi; \text{$\eta_1$}=0; \text{$\eta_2$}=\frac{\pi }{2}; \theta =\frac{\pi }{2}$ or  $\text{$\Delta \beta$}=2 \pi ; \text{$\eta_1 $}=0;\text{$\eta_2$}=\frac{\pi }{2};\theta =\frac{\pi }{2}$.}
With the above conditions, we get the following expression:
\begin{equation}\label{df3}
\mathcal{F}  =   1-\frac{\left(2\gamma-\gamma^2 \right)}{4} \big(3-\cos \Delta \beta\big).
\end{equation}
Comparing the  expressions of $\mathcal{F}$ obtained in Eqs. (\ref{dc2}), (\ref{ac3}) and (\ref
{df3}), we find that the \(\gamma\)-dependence of the above equation is similar with DC. 
It  can also be confirmed from Figs. \ref{fig2} and \ref{dp3}.  

For fixed values of \(\Delta \delta=0\), \(\theta= \pi/2\) and \(\gamma=1/2\), we find
\begin{align}\label{df4}
\mathcal{F} =
\frac{1}{16} \big(7 + 3 \cos\Delta \eta + 3(1 + \cos\Delta \eta)\cos\Delta \beta\big).
\end{align}
The trends of \(\mathcal{F}\) is again similar to the one obtained for DC (see Figs. \ref{dp4})  and \ref{fig5}).




\textit{Fixing the control parameters to obtain complementarity above a certain threshold}: Given a fixed amount of noise, $\gamma$, in a noisy channel, one can ask the following: To obtain $\mathcal{F}\geq \mathcal{F}_{threshold}$, how can we control  the parameters involved in unitaries? We will now show that this is indeed possible. To illustrate, let us demand $\mathcal{F}\geq 0.9$.  In the following table,  we find the control parameters $\Delta\eta$, $\Delta\beta$ and $\Delta\delta$  which lead to $\mathcal{F}\geq 0.9$ for all noisy channels irrespective of the choices of \(\theta\).  Note that \(\mathcal{F}\) is independent of phase, \(\phi\), of the quantum states. 
\begin{table}[h!]
	\centering
	\begin{tabular}{ |c|m{2.2cm}|m{2.2cm}|m{2.2cm}| } 
		\hline
		Channels &DC ($\gamma=0.5$)& ADC ($\gamma=0.07$) & PDC ($\gamma=0.07$)\\ 
		\hline
		$\Delta \eta$&$0.5$ & $\pi$ & $\pi$ \\
		\hline 
		$\Delta \beta$&$0.2$ & 0 & 0 \\ 
		\hline
		$\Delta \delta$&0 & 0 & 0 \\ 
		\hline
	\end{tabular} 
\caption{The above table gives a sample values of the control parameters for which $\mathcal{F}\geq 0.9$ independent of the parameters $\theta$ and $\phi$. These values have been obtained from the cases considered in figures (1)\ref{fig5} for depolarizing channel (2) \ref{adc1} for amplitude damping channel and (3) \ref{dp1} for phase damping channel.}

\end{table}

Notice that for symmetric quanton states, the complementarity relation is independent of state parameters $\theta$ and $\phi$ in case of depolarizing channel  and hence it is easier to find the values of control parameters in this case compared to ADC and PDC.


\section{Conclusion}
\label{sec_conclu}

Wave-particle duality was known as the most impressing demonstration of the nonclassical feature present in quantum theory from the very beginning of its origin. This characteristics of the quantum system have been experimentally verified in various kinds of interferometric set-up and are presently known as interferometric complementarity. In recent years, the new duality relation is proposed involving coherence and path distinguishability of quanton. Here we consider the picture where the detectors are influenced by the noisy environment, thereby affecting both coherence as well as distinguishability and hence complementarity relation. It is found that there exist certain initial detector states and parameters specifying interactions between the detector and the quanton such that the decohering effect of noisy detectors can be reduced and even in some cases,  is possible to completely wash out. In this scenario, relevant tuning parameters are identified and studied extensively. Specifically, the asymmetric initial quantum states are found to be more advantageous concerning the saturation of the complementarity than symmetric quantum states, irrespective of the noise models for some choices of parameters. Our investigations shed light on the consequence of decoherence in the experimental test of wave-particle duality. It can be interesting to study other factors like memory and non-Markovian channels which can influence the duality relation.


\section{Acknowledgement}
The research of GS was supported in part by the INFOSYS scholarship for senior students.

\appendix

\section{Quantum noise }\label{appen}

Let us discuss briefly the noisy channels considered in this paper. We will fix the notations here.  

\subsection{Depolarizing channel}
\label{subsec:depolK}
The quantum noise produced by a depolarizing channel takes a single  qubit state, $\rho$ to a maximally  mixed state $I/2$ with  probability $\gamma$ and the state is left intact with the rest of the probability. Therefore, we have
\begin{equation}
\label{21}
\rho \longrightarrow \mathcal{E_{D}}(\rho)= \frac{\gamma I}{2}  +(1-\gamma) \rho.
\end{equation}
In the operator-sum representation, Eq. (\ref{21}) can be represented as
\begin{equation}
\label{21a}
\mathcal{E_{D}}(\rho)= \left(1-\frac{3\gamma}{4}\right) \rho  + \frac{\gamma}{4} \sum_{i=1}^{3} \sigma_{i} \rho \sigma_{i}^\dagger,
\end{equation}
where $\sigma_i$ are Pauli matrices.

\subsection{Amplitude damping channel}
\label{subsec:ampdampK}

The amplitude damping channel (ADC) is a schematic model that
describes the interaction of a two-level atom with the electromagnetic field (environment). 
The evolution in this case is described by a unitary transformation that acts on atom and environment which  can  always be taken in a pure state without any loss of generality. The corresponding transformation is  given by
\begin{eqnarray}
\label{12}
\left| 0\right\rangle _{S}\left| 0\right\rangle _{E} & \rightarrow  & \left| 0\right\rangle _{S}\left| 0\right\rangle _{E} \\
\left| 1\right\rangle _{S}\left| 0\right\rangle _{E} & \rightarrow  & \sqrt{1-\gamma}\left| 1\right\rangle _{S}\left| 0\right\rangle _{E}+\sqrt{\gamma}\left| 0\right\rangle _{S}\left| 1\right\rangle _{E}
\end{eqnarray}
where the parameter \( \gamma \) denotes the dissipation strength.
Physically, it tells us that if an atom was in an excited state, \( \left| 1\right\rangle _{S} \),  it makes a transition to the ground state \( \left| 0\right\rangle _{S} \)  with probability \( \gamma \) by emitting  a photon. The environment as a result makes a transition
from ``no-photon'' state \( \left| 0\right\rangle _{E} \) to the ``one-photon'' state \( \left| 1\right\rangle _{E} \).
Tracing out the environment, one obtains the Kraus operators, \( K_{i} \), given by
\begin{equation}
\label{13}
K_{1}=\left( \begin{array}{cc}
1 & 0\\
0 & \sqrt{1-\gamma}
\end{array}\right),\, K_{2}=\left( \begin{array}{cc}
0 & \sqrt{\gamma}\\
0 & 0
\end{array}\right). 
\end{equation}
The initial state \(\rho\) changes in presence of ADC can be written as
\begin{equation}
\label{15}
\mathcal{E_{AD}}(\rho) =K_{1}\rho K^{\dagger }_{1} + K_{2}\rho K^{\dagger }_{2}.
\end{equation}

\subsection{Phase damping channel}
\label{subsec:phasedampK}

Phase damping is a unique quantum mechanical noise process, which explains the loss of quantum information without the loss of energy. The effect of noise can be given by following Kraus operators, \( K_{i} \): 
\begin{equation}
\label{22}
\begin{array}{l}
{K_1} = \sqrt{1-\gamma} \, I, \\
{K_2} = \sqrt{\gamma} \left( {\begin{array}{*{20}{l}}
	1&0\\
	0&0
	\end{array}} \right),\\
{K_3} = \sqrt {\gamma} \left( {\begin{array}{*{20}{l}}
	0&0\\
	0&1
	\end{array}} \right).
\end{array} 
\end{equation}
The state  evolves in this case as
\begin{equation}
\label{23}
\mathcal{E_{PD}}( \rho) =  K_{1}\rho K^{\dagger }_{1}+ K_{2}\rho K^{\dagger }_{2}+ K_{3}.\rho. K^{\dagger }_{3}
\end{equation}

\end{document}